\documentclass[twocolumn,aps,superscriptaddress]{revtex4}
\usepackage[utf8x]{inputenc}
\usepackage{ucs}
\usepackage{amsmath}
\usepackage{amsfonts}
\usepackage{amssymb}
\usepackage{makeidx}
\usepackage{graphicx}
\usepackage{xcolor}
\usepackage{listings}
\usepackage{subfig}
\usepackage[sort&compress]{natbib}

\graphicspath{ {figs/} }

\begin{document}
\author{M. H. Jensen} \affiliation{Glass \& Time, IMFUFA, Department
  of Science and Environment, Roskilde University, P.O. Box 260,
  DK-4000 Roskilde, Denmark} \affiliation{Laboratoire Léon Brillouin,
  CNRS CEA -UMR 12, CEA Saclay, 91191 Gif-sur-Yvette
  Cedex, France}

\author{C. Gainaru} 
\affiliation{Fakultät Physik, Technische Universität Dortmund, 44221 Dortmund, Germany}

\author{C. Alba-Simionesco} \affiliation{Laboratoire Léon Brillouin,
  CNRS CEA -UMR 12, CEA Saclay, 91191 Gif-sur-Yvette Cedex, France}

\author{T. Hecksher} 
\affiliation{Glass \& Time, IMFUFA, Department
  of Science and Environment, Roskilde University, P.O. Box 260,
  DK-4000 Roskilde, Denmark}

\author{K. Niss} 
\affiliation{Glass \& Time, IMFUFA, Department
  of Science and Environment, Roskilde University, P.O. Box 260,
  DK-4000 Roskilde, Denmark}

\title{Slow rheological mode in glycerol and glycerol-water mixtures}

\date{\today}

\begin{abstract}
  \begin{minipage}{0.4\linewidth}
  Glycerol-water mixtures were studied at molar concentrations ranging
  from $x_\text{gly} = 1$ (pure glycerol) to $x_\text{gly}=0.3$ using
  shear mechanical spectroscopy. We observed a low frequency mode in
  neat glycerol, similar to what is usually reported for monohydroxy
  alcohols. This mode has no dielectric counterpart and disappears
  with increased water concentration. We propose that the
  hydrogen-bonded network formed between glycerol molecules is
  responsible for the observed slow mode and that water acts as a
  plasticizer for the overall dynamics and as a 
  lubricant softening the hydrogen-bonding contribution to the
  macroscopic viscosity of this binary system.
  \end{minipage}
  \hspace{0.5cm}
  \begin{minipage}{0.4\linewidth}
  \includegraphics[scale=0.7]{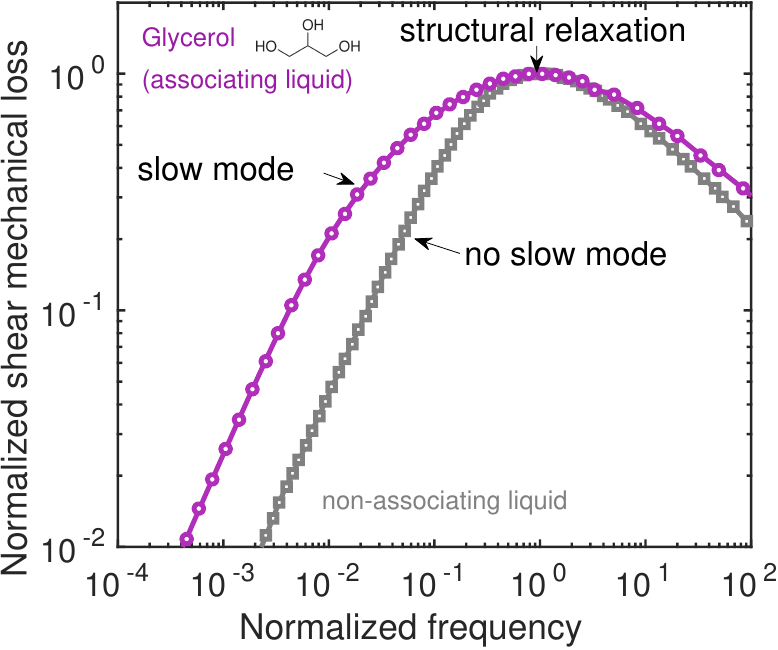}  
  \end{minipage}
  
\end{abstract}

\maketitle
\section{Introduction}
Hydrogen bonds play an important role in biology as well as
technology. They govern the chemical and physical properties of bulk
water, alcohols and sugars
\cite{Pauling1939,Jeffrey1991,Desiraju1999}, they play a central role
in protein and DNA structures as well as being central in water
interactions in other materials \cite{Waterbook2007}. For this reason
there is intense scientific focus on understanding the hydrogen
bonds. Glycerol (propane-1,2,3-triol) is a small molecule with three
hydroxyl groups which supports the forming of a 3D hydrogen-bonded
network penetrating the bulk liquid. It is the most used liquid in
fundamental studies of glassy dynamics, e.g.  Refs.\
\cite{Gibson1923,Wang1971,Schiener1996,Berthier2005,Pezeril2009,Albert2016},
as well as being widely used in technology, notably due to its
cryoprotectant properties \cite{Salt1958,Dashnau2006,Li2008} which are
intimately connected to its large affinity to form hydrogen bonds. In
this respect it is important to gain an understanding of the
interaction between the glycerol hydrogen-bonded network and water not
only from a fundamental point of view but also for unraveling the
mechanisms behind the antifreezing effect of glycerol \cite{Popov2015,
  Murata2012}.

Hydrogen-bonded molecules inherently comprise atoms with large
electronegativity contrast and thus exhibit large electric
polarizability, meaning that they have a strong signal in dielectric
spectroscopy (DS). DS has therefore been widely used to study the
dynamics of hydrogen-bonded liquids particularly in the supercooled
regime. The dielectric properties of glycerol have been studied in an
extremely broad dynamic range \cite{Lunkenheimer2000} and from its
boiling point down to cryogenic temperatures \cite{Gainaru2005}. In
the supercooled regime its dielectric loss spectra are dominated by
the structural relaxation process which can be fitted very well by a
single-peak function. In this respect glycerol resembles the behavior
of non-associated, van der Waals liquids. Glycerol-water mixtures are
also extensively studied with DS \citep{Sudo2002, Puzenko2005,
  Puzenko2007, Hayashi2005, Hayashi2006, Popov2015} and also in this
case the dielectric signal show no obvious signature of the hydrogen
bonding association.

On the other hand, many alcohols with a single hydroxyl group per
molecule (monoalcohols) display a much more complex dielectric pattern
including an absorbtion mode which is slower than the structural
relaxation, the so-called Debye process
\cite{Murthy1993,Huth2007,Bohmer2014}. Currently there is a widespread
contention that the Debye dynamics in monoalcohols is related to the
formation of hydrogen-bonded supramolecular structures of polar short
chains which for certain systems be converted to non-polar rings at
temperatures well above $T_g$ as illustrated by the temperature
dependence of the Kirkwood correlation factor \cite{Singh2013}.  The
hydrogen-bonded structures can also give rise to an additional peak in
the static structure factor at low Q-vectors
\cite{Bohmer2014,Singh2015}, although the reverse is not true, a
prepeak in the structure is not systematically associated with a
slower than structural relaxation feature in the dynamics
\cite{Morineau1998, Mandanici2005}. The family of liquids displaying a
dielectric Debye process \cite{Gao2013} includes, besides
monoalcohols, secondary amides \cite{Wang2005}, protic ionic liquids
\cite{Cosby2015}, and several pharmaceuticals
\cite{Adrjanowicz2013,Rams-Baron2015}.

The Debye process was for a long time considered merely a dielectric
feature \cite{Pawlus2013}, however recent rheology shear
\cite{Gainaru2014} and bulk modulus \cite{Hecksher2016} experiments
studies have demonstrated otherwise: Low molecular weight monoalcohols
exhibit slower-than-structural-relaxation dynamics otherwise
considered specific to covalently bonded structures. For polymeric
melts it is well established that the entropic elasticity of
covalently bonded structures leads to the emergence of slow mechanical
modes which govern their steady-state viscosity and implicitly their
macroscopic flow \cite{Ferry1980}.

Such polymer-like mechanical dynamics was mainly discussed in
literature for monoalcohols, but more recent investigations have also
discovered it in an amine system \cite{Adrjanowicz2015}. Important for
the present context is that although amine systems are also able to
sustain hydrogen bonds, the slow rheology process was found to not
have a dielectric counterpart, contrary to the situation for
monoalcohols. The systematic of the separation between Debye and alpha
process seen in dielectrics is apparently conserved in the rheological
signal for some systems \cite{Gainaru2014,Hecksher2016} while the
correlation is less clear in a study of the $N$-methyl-3-heptanol
series ($N = 2, ...\,, 6$) \cite{Hecksher2014}. These observations
suggest that mechanical spectroscopy is a important complementary
technique which can offer a different perspective, thus shed new light
in the dynamical signatures of hydrogen-bonded structures in liquids.

To our knowledge this type of slow mechanical mode has not so far been
identified in polyalcohols.  In this paper we present data on glycerol
and glycerol rich glycerol-water mixtures obtained with broadband
shear mechanical spectroscopy revealing a slow contribution to the
spectral density, which we ascribe to the dynamics of the
hydrogen-bonded network. The strength of this slow mode decreases with
increasing the water content and virtually disappears at equimolar
composition, $x_\text{gly}=0.5$, corresponding to 20 volume percent of
water. This indicates that water significantly changes the properties
of the glycerol hydrogen-bonded network. Our data demonstrate that the
shear mechanical spectra are able to reveal aspects of the relaxations
dynamics in hydrogen-bonded materials which give no contribution to
the dielectric signal.

\section{Experiments}
The glycerol-water mixtures were prepared using water from an
Arium\textregistered 611 ultrapure water system and glycerol from
Sigma Aldrichs at $99.9\%>$ purity. The concentrations used were
$x_\text{gly} = 1,\ 0.8,\ 0.7\ 0.6\ 0.5$ and $0.3$. Table
\ref{tab:overview} gives an overview of the concentrations used. For
the measurement on neat glycerol the sample was loaded into the cell
under nitrogen atmosphere in order to avoid uptake of water from the
air.

\begin{table}[ht!]
\centering
\caption{Overview of the concentrations used.}\label{tab:overview}	
\begin{tabular}{ll|ccc}
  $m_{\text{glycerol}}$ & $m_{\text{water}}$ &
  $x_\text{gly}^\text{mol}$ & $x_\text{gly}^\text{vol}$ &
  $x_\text{gly}^\text{weight}$ \\ 
  \hline
  6.3183 & 0 & 1 & 1 & 1 \\
  6.3152 & 0.3081 & 0.7999 & 0.9419 & 0.9533 \\
  6.3045 & 0.5282 & 0.7004 & 0.9046 & 0.9228 \\
  6.3082 & 0.8225 & 0.6000 & 0.8589 & 0.8846 \\
  6.3230 & 1.2324 & 0.5002 & 0.8024 & 0.8365 \\
  6.3388 & 2.8766 & 0.3012 & 0.6362 & 0.6878 \\
  \hline
\end{tabular}
\end{table}

The employed experimental method is the “piezoelectric shear-modulus
gauge” (PSG) technique \citep{Christensen1995} covering the frequency
range from mHz to kHz corresponding to timescales ranging from ca. 100
seconds to ca. 10 microseconds. The measured quantity is the complex
frequency-dependent shear modulus $G^*(\nu) = G'(\nu)+iG''(\nu)$
defined for a harmonic oscillation stress and strain by
\begin{equation}
  \label{eq:1}
  G^*=\frac{\sigma_0}{\gamma_0} \text{e} ^{i\delta}
\end{equation}
where $\sigma_0$ and $\gamma_0$ are the amplitudes of the stress and
strain respectively and $\delta $ is the phase difference between
stress and strain.  The complex modulus will have an imaginary part
whenever $\delta \neq 0$.  Data were measured upon cooling in the
range from roughly 30 degrees above $T_g$ down to $T_g$ in steps of a
couple of degrees.

\section{Results and Analysis}

\subsection{Raw data and temperature dependence}
\begin{figure}[h!]
\includegraphics[width=8cm]{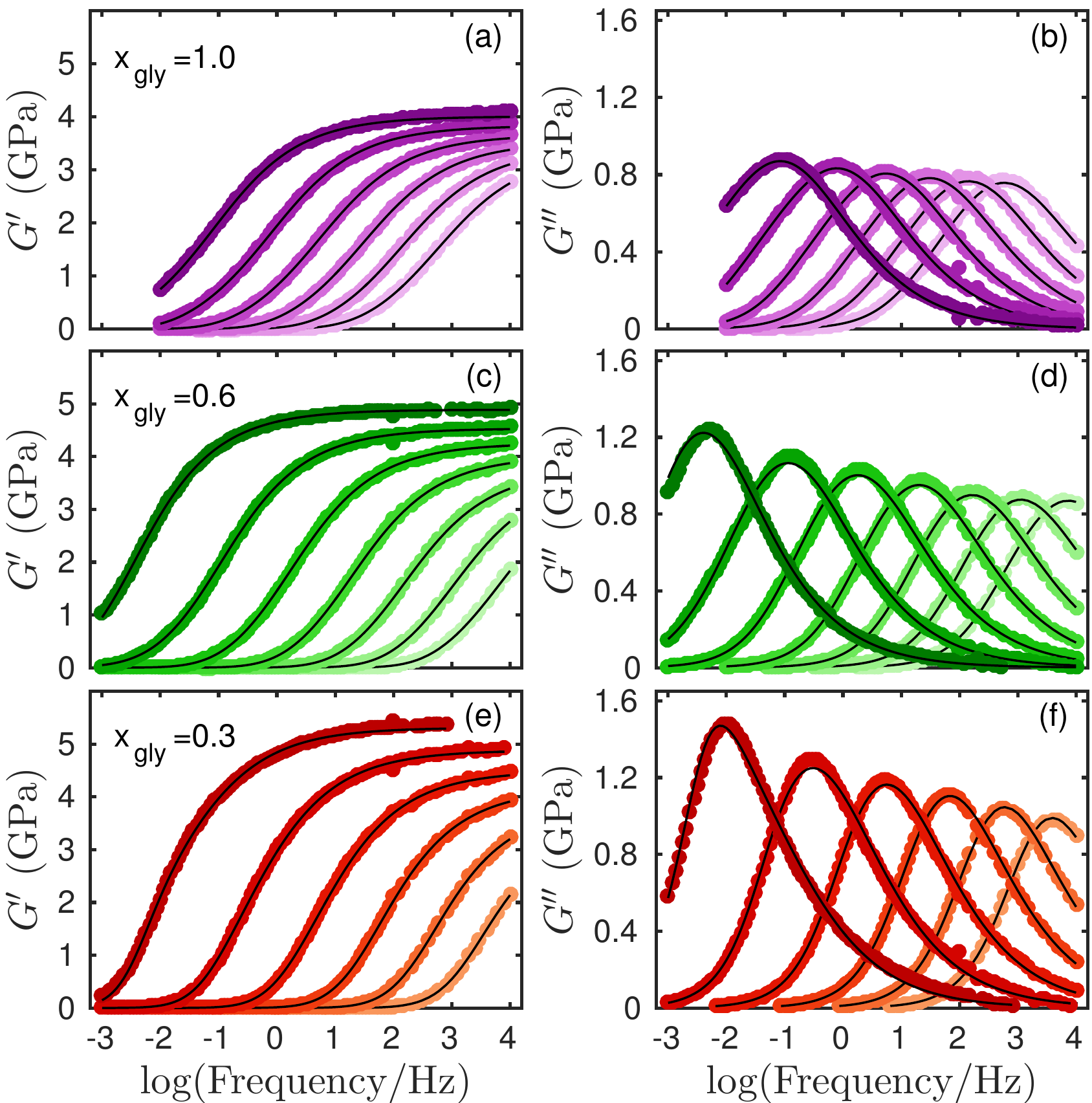}
\caption{\label{fig:dataWithFits}The complex shear modulus for three
  of the six concentrations studied. Temperatures shown for
  $X_{\text{gly}}=1$ range from 190~K to 210~K in step of 4~K, for
  $X_{\text{gly}}=0.6$ from 175~K to 205~K in step of 5~K, and for
  $X_{\text{gly}}=0.3$ from 165~K to 190~K in step of 5~K.  The black
  lines are fits of the data to an extended Maxwell model presented in
  Ref.\ \cite{Hecksher2017}. An example of the fits on a log-log 
scale  is shown in Fig.\ \ref{fig:tauMtauAlphaDecoupling}(c).}
\end{figure}

In Fig.\ \ref{fig:dataWithFits} we show the frequency dependence of
the real and imaginary part of the shear modulus plotted on linear
$y$-scales, for neat glycerol and two out of the five investigated
mixtures. The real part of the shear modulus goes to zero at low
frequencies indicative of the viscous flow, while it reaches a finite
value at high frequencies, because the liquid behaves like a solid at
short time scales. The transition between the viscous and the elastic
behaviors gives rise to a peak in the imaginary part, commonly
referred to as the loss peak. The loss peak moves to lower frequencies
as the liquid is cooled and the solid-like limiting behavior will fill
the entire frequency range when the liquid reaches the glass
transition temperature.

\begin{figure}[h]
  \includegraphics[width=8cm]{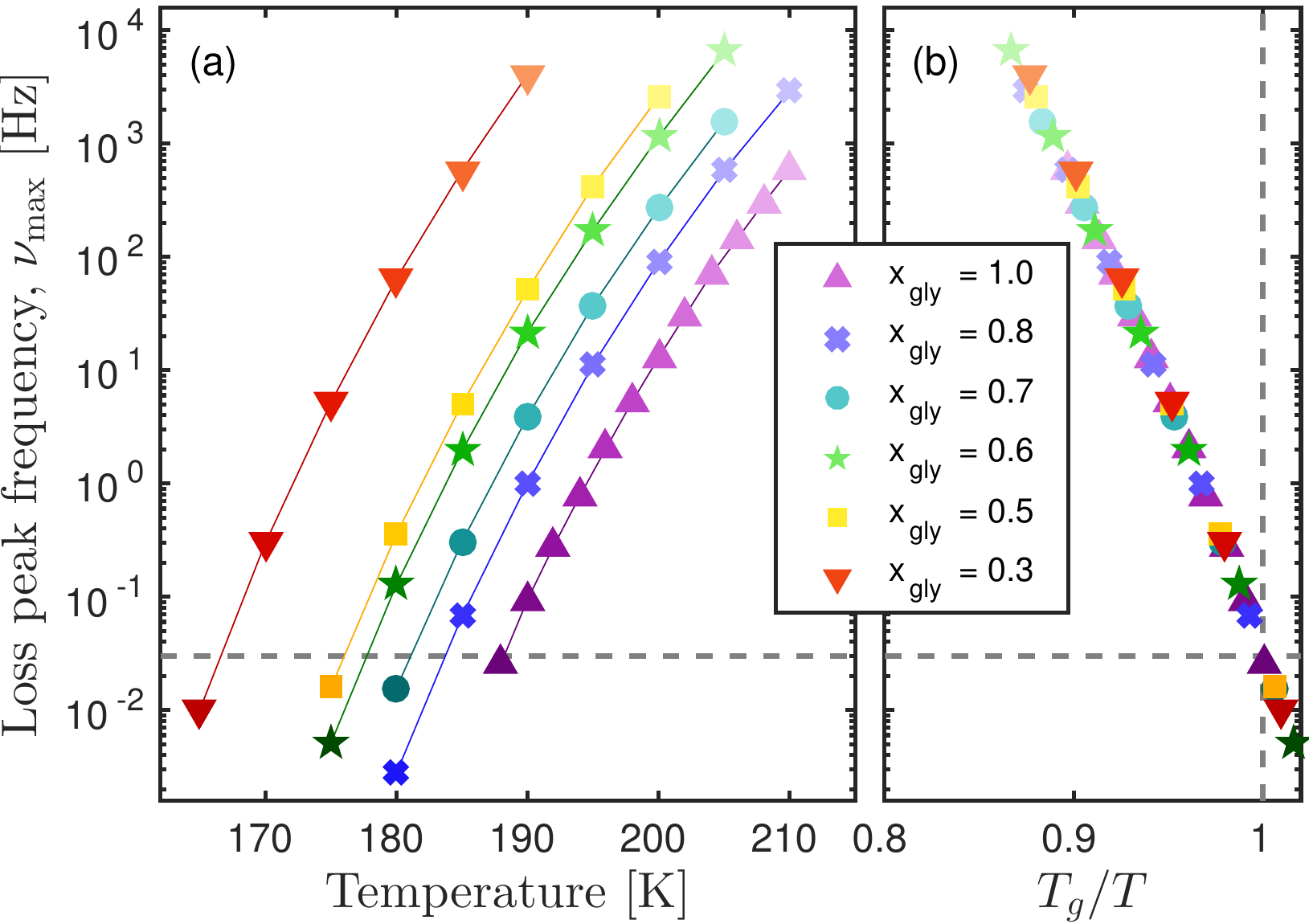}
  \caption{Relaxation map for the investigated glycerol-water
    mixtures. (a) Loss peak frequencies as a function of temperature
    for the different concentrations. The plasticising effect of water
    lowers $T_g$ and shifts the curves to lower temperature with
    increasing water content. (b) All curves collapse when plotted
    against inverse temperature scaled with $T_g$ (defined as the
    $\nu_{\text{max}}(T_g) = 30$~mHz and shown by the dashed
    line). Thus the temperature behavior of the relaxation time is
    surprisingly similar across the concentration range explored
    here. \label{fig:tauVTemp}}
\end{figure}
\begin{figure}[h!]
  \includegraphics[width=7cm]{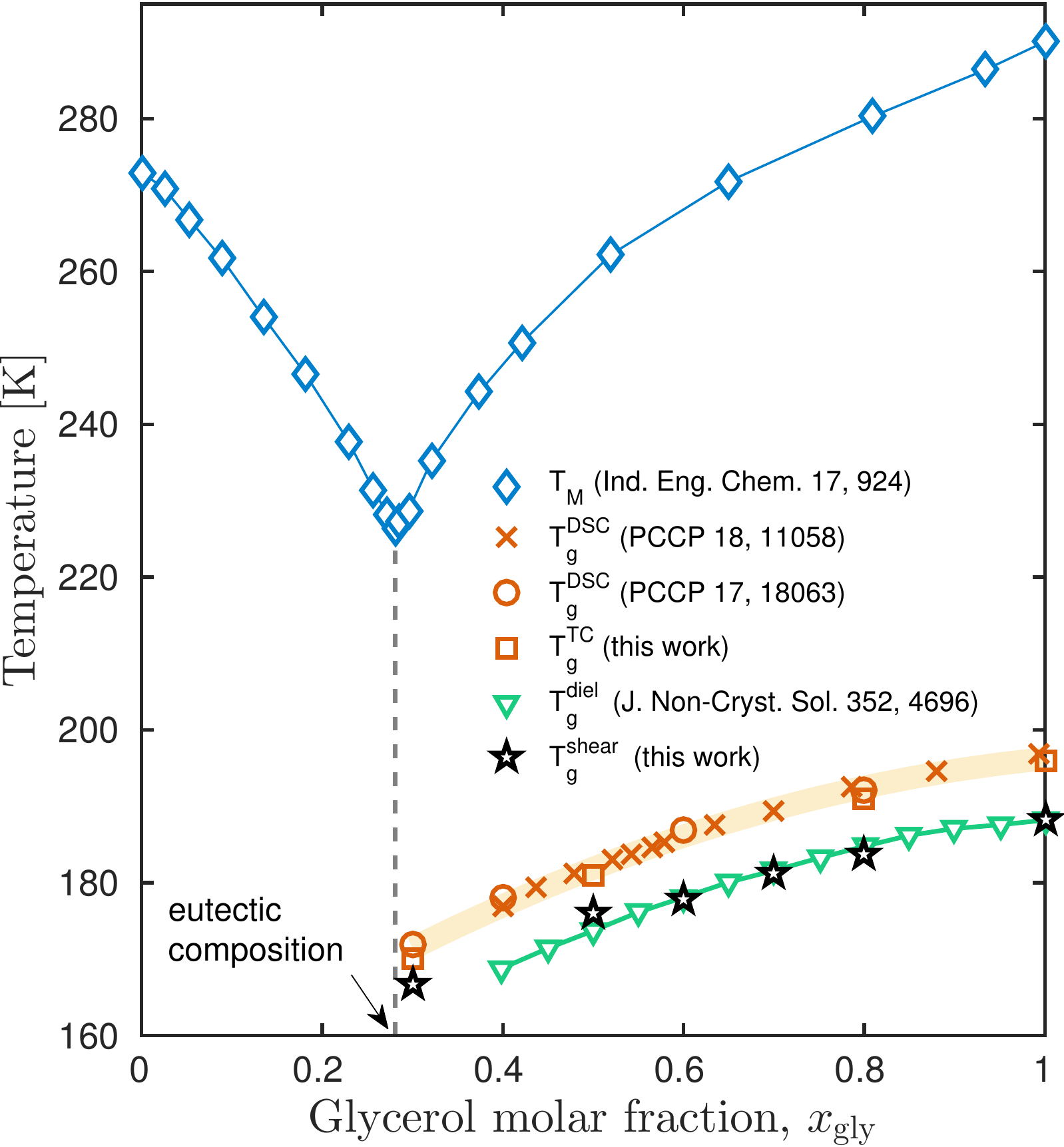}
  \caption{The glass and melting transition temperatures as a function
    of glycerol concentration. The eutectic point occurs at
    $X_{\rm{gly}}=0.28$ (melting point data from
    Ref.\ \cite{Lane1925}). Glass transition temperatures roughly obeys
    the 2/3 rule in the range of concentrations explored in this work
    (all above the eutectic composition). The $T_g$ curves depend
    slightly on the method of determination: calorimetric curves --
    differential scanning calorimetry (DSC) and thermalization
    calorimetry (TC) -- are higher than the dielectric and shear, but
    follow the same trend. DSC data are from Ref.\ \cite{Bachler2016}
    and \cite{Popov2015}. Dielectric data are found in Ref.\
    \cite{Hayashi2005}, where the $T_g$ is defined as $\tau(T_g)
    \equiv 100$~s. Shear data are from this work using $T_g$ defined
    in Fig.\ \ref{fig:tauVTemp}. TC data are obtained by the method
    described in Ref.\ \cite{Jakobsen2016}. \label{fig:TgvsXg}}
\end{figure}

In Fig.\ \ref{fig:tauVTemp}(a) we show the evolution of the frequency
of the loss peak maximum, $\nu_{\text{max}}$, with temperature. In
this representation it becomes clear that adding water to gly\-cerol
has a plasticizing effect, as for a given temperature it increases the
loss peak frequency. This behavior is consistent with the dielectric
measurements \citep{Puzenko2005,Popov2015}. The‭ ‬water-induced increase
in the overall molecular mobility‭ ‬render‭ a lower glass transition
temperature,‭ ‬$T_g$,‭ ‬in harmony with previous results obtained from
differential scanning calorimetry performed on the same mixtures
\cite{GlyProp1963}. ‬In the following we will introduce‭
‬$T_{g}^\text{shear}$‭ ‬as the temperature at which the modulus loss peak
frequency‭ $\nu_{\text{max}}$‭ ‬reaches‭ ‬30‭~‬mHz.‭ This frequency is chosen
to avoid extrapolation and corresponds to a relaxation time of
5~seconds. ‬By such means one is able to compare the current results
with those previously obtained via complementary techniques,‭ ‬as done
in Fig.‭ ‬\ref{fig:TgvsXg}.‭ ‬Here $T_g$-data from other methods are added
to the phase diagram of the glycerol-water mixture.

In Fig.‭ ‬\ref{fig:TgvsXg}‭ ‬one may note that the lowest glycerol
concentration considered in the present study is close to the eutectic
point of this mixture,‭ ‬which occurs at a glycerol molar concentration
of‭ $x_\text{gly} = 0.28$.  In the investigated concentration range the
decrease in‭ ‬$T_g$‭ ‬occurring upon lowering $x_\text{gly}$ mirrors the
depression of‭ ‬melting temperature, $T_M$,‭ ‬as the two characteristic
temperatures comply‭ ‬well with the empiric proportionality $T_g/T_M
\approx 2/3$ \cite{Donth2001}.

Returning now to Fig.\ \ref{fig:tauVTemp}, panel (b) depicts the loss
peak frequencies on a reduced temperature scale. In particular, for
each concentration the inverse temperature axis is multiplied by the
corresponding value of $T_{g}^{\textrm{shear}}$. One can easily observe that such
a scaling render a good overlap of all the data sets, disregarding the
water concentration. In other words, the relative temperature
dependence of the relaxation time, which can be quantified by the
fragility $m=d \log \tau / d(T_g/T)$ \cite{Angell1991}, is
surprisingly unaffected by the addition of water.

\subsection{Spectral shape and slow mode}
In order to study the variation of the spectral shape with water
content we plot the data from Fig.\ \ref{fig:dataWithFits} on a
logarithmic scale, and normalize with respect to both the loss peak
frequency, $\nu_{\text{max}}$, and the amplitude of the signal.

\begin{figure}[h!]
  \includegraphics[width=8cm]{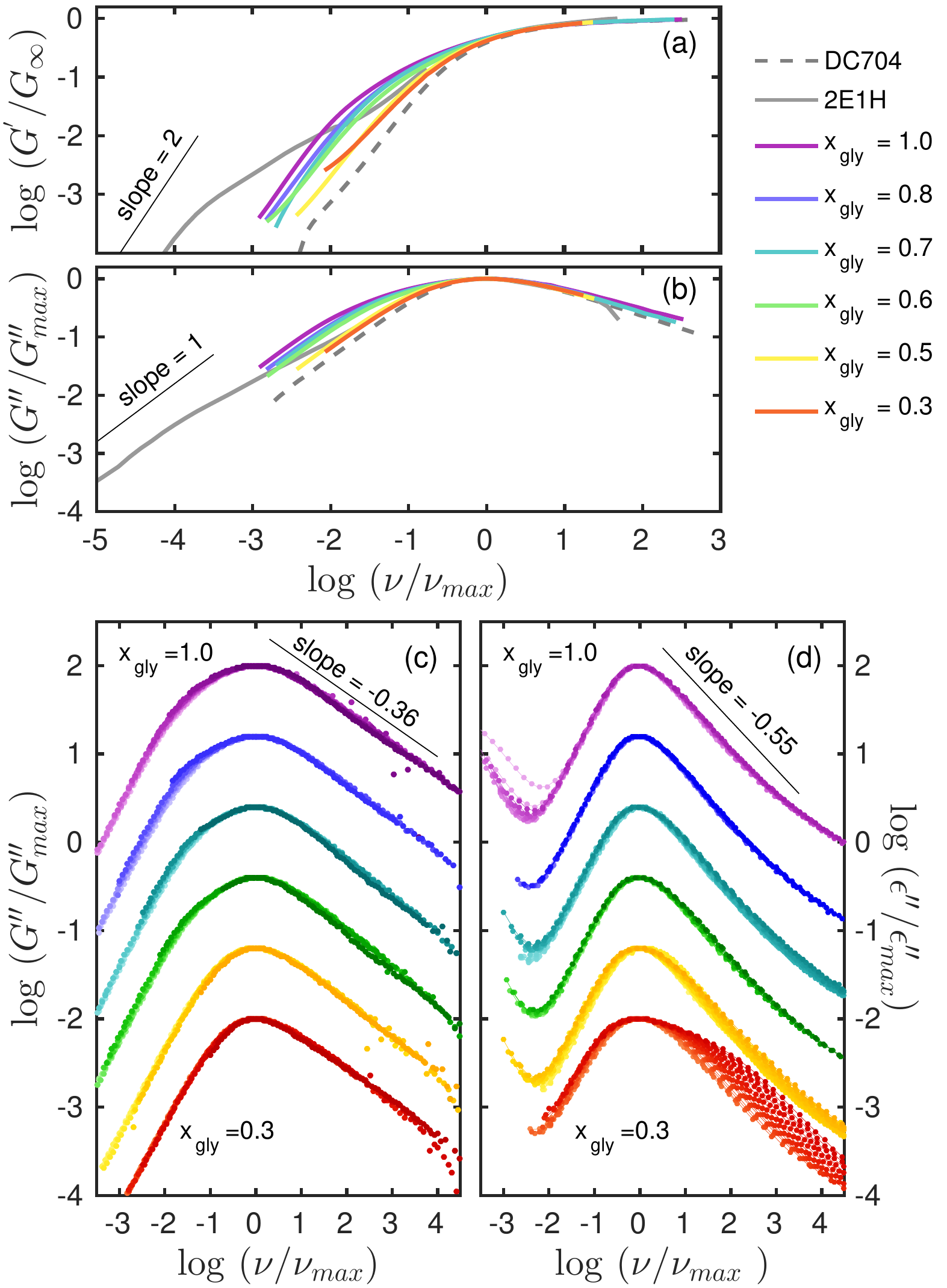}
  \caption{\label{fig:shoulder}The spectral shape of glycerol and its
    mixtures with water. (a)+(b) Real and imaginary part of the
    normalized shear modulus of a single spectrum for each
    concentration.  For comparison we have added the liquids
    2-ethyl-1-hexanol (2E1H) and tetramethyl-tetraphenyl-trisiloxane
    (DC704) (data from Ref. \cite{Gainaru2014} and
    Ref. \cite{Jakobsen2005}). (c)+(d) Normalized TTS plots of the
    loss peak of all the shear spectra (this work) and the dielectric
    spectra (data from Ref. \citep{Popov2015}).  Each concentration
    has been given an offset to make comparison between the
    concentrations easier. Clearly, the spectral shapes in shear and
    dielectric are quite different (see also Fig.\
    \ref{fig:sheardiel}) as is their evolution with concentration.}
\end{figure}

This is done in Fig.\ \ref{fig:shoulder}(a) and \ref{fig:shoulder}(b)
which contain for clarity reasons a single spectrum for each
investigated concentration. The corresponding temperature was chosen
so that the relaxation time of the different materials roughly
coincide.  All spectra clearly exhibit at lowest frequencies the
$G''(\nu) \propto \nu$ and $G'(\nu) \propto \nu^2$ power laws which
are characteristic for viscous flow \cite{Harrison1976}.  However, for
the neat glycerol as well as for the mixtures with highest glycerol
concentrations a low frequency shoulder can be identified in both the
real and the imaginary parts before the viscous flow sets in. This
spectral contribution which is slower than the structural (alpha)
relaxation strongly resembles the mechanical mode which is discussed
for monoalcohols in connection with the formation of
hydrogen-bonded supramolecular structures.

To further strengthen this point, we have included for comparison the
liquids 2-ethyl-1-hexanol (2E1H) and
tetramethyl-tetraphenyl-trisiloxane (DC704) to Fig.\
\ref{fig:shoulder}(a) and \ref{fig:shoulder}(b). 2E1H is a
paradigmatic monoalcohol which is known to exhibit a supra-molecular
relaxation clearly seen as distinct slow mode in the mechanical
spectra as well as a strong Debye signal in the dielectric spectra
\cite{Gainaru2014}, while DC704 is a van der Waals liquid with no
additional modes and a ``simple'' behavior in general
\cite{Jakobsen2005,Roed2015}.  As observed in Fig.\
\ref{fig:shoulder}(a) and \ref{fig:shoulder}(b) the shape of the
mechanical spectra of the glycerol-water mixtures fall between the
spectra of these two liquids, with neat glycerol being closer to 2E1H,
while the glycerol-water mixtures resemble more DC704 with the
increase of water content. Accordingly, neat glycerol behaves like a
liquid with supramolecular structure dynamics while adding water makes
it behave more like a liquid devoid of supramolecular associations.

In Fig.\ \ref{fig:shoulder}(c) and \ref{fig:shoulder}(d) we show the
normalized imaginary part of all the shear spectra (this work) and the
corresponding dielectric loss spectra (data from
Ref. \citep{Popov2015}).  The scaled data are presented in arbitrary
units, as for each concentration has been given an offset to make
comparison of the concentrations easier. Clearly no low-frequency
Debye mode is visible in the dielectric spectrum of glycerol nor any
of the studied glycerol-water mixtures. Thus, whatever mechanism is
responsible for the observed slow rheological mode is not giving rise
to a net electric polarization. This is markedly different from the
situation for monoalcohols where a strong dielectric signal emerges
due to collective dynamics of the individual (molecular) dipoles
\cite{Bohmer2014}.

In order to quantify the degree of separation between the mechanically
active slow mode and the alpha relaxation we introduce the classical
Maxwell time given by:
\begin{equation}\label{eq:tauMaxwell}
  \tau_M =  \frac{\eta_0}{G_\infty}
\end{equation}
where $\eta_0$ and $G_\infty$ are given by the limiting behaviors:
\begin{eqnarray}
  \eta_0 \equiv \lim\limits_{\omega \rightarrow 0}(G''(\omega)/\omega)\\
  G_\infty \equiv \lim\limits_{\omega \rightarrow \infty}(G'(\omega))\,.
\end{eqnarray}

\begin{figure}[h!]
  \begin{center}
    \includegraphics[width=8.2cm]{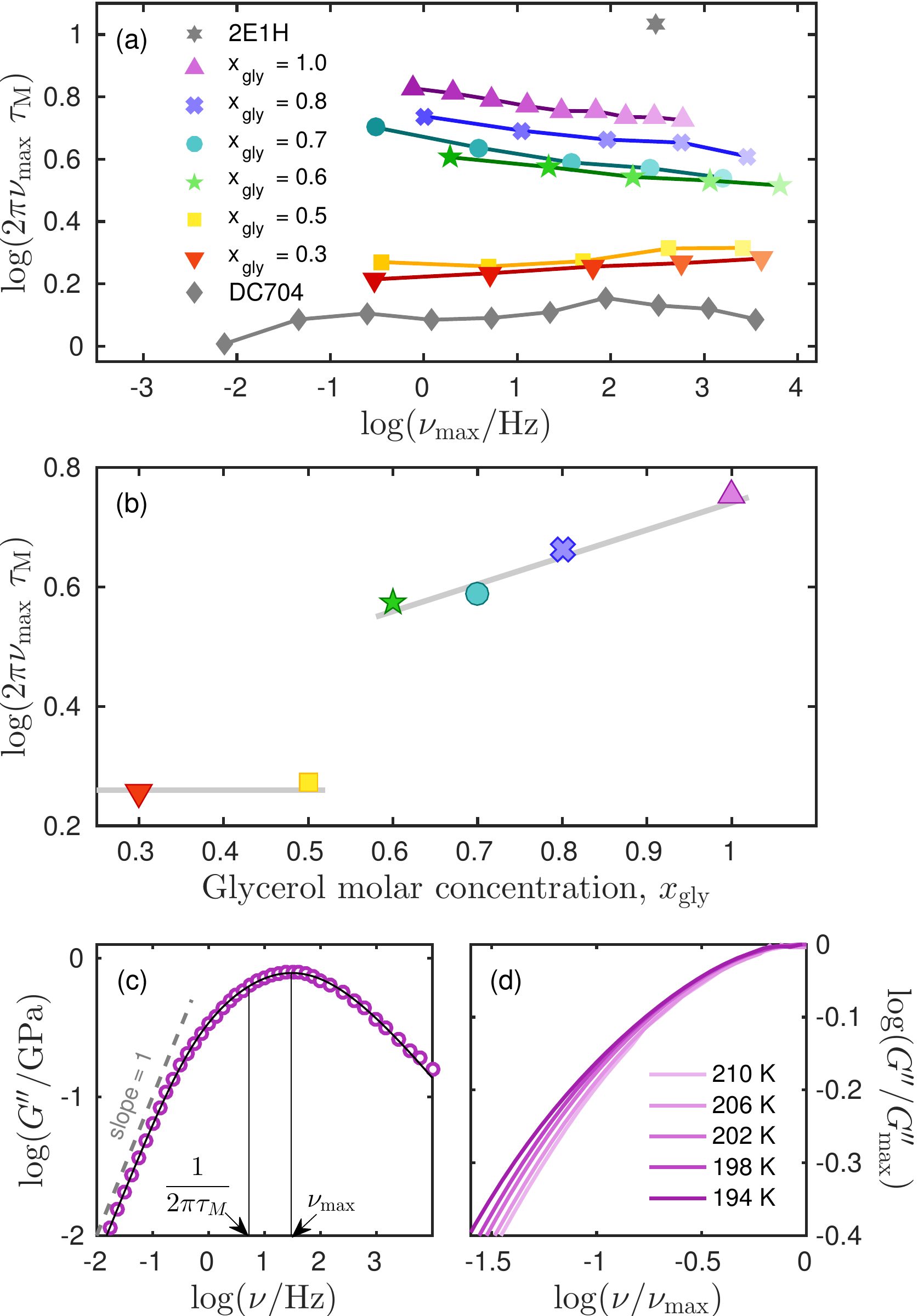}
  \end{center}
  \caption{\label{fig:tauMtauAlphaDecoupling}Decoupling between the
    two timescales $1/(2\pi\nu_{\text{max}})$ and $\tau_M$. (a)
    Decoupling index as a function of loss peak frequency. For
    $x_\text{gly} \geq 0.6$ the decoupling index decreases slightly
    with temperature. For $x_\text{gly} \leq 0.5$ it is constant or
    perhaps slightly increasing. For comparison the decoupling index
    of DC704 and 2E1H is shown. (b) The decoupling index evaluated at
    a loss peak frequency close to 100~Hz as a function of glycerol
    concentration. The index is increasing as a function of
    concentration with a dramatic change between $x_{\rm{gly}}=0.5$
    and $x_{\rm{gly}}=0.6$. Lines are guides to the eye that highlight
    the different trends in the two regions. (c) Illustrating for a
    single neat glycerol spectrum the difference between the loss peak
    frequency and the fitted Maxwell time. Circles are data points,
    the full line is the fit. (d) Zoom of the neat glycerol curves in
    Fig.\ \ref{fig:shoulder} showing the deviation from TTS.}
\end{figure}

In the simplest case, corresponding to exponential relaxation in the
time domain, one has $2 \pi \nu_{\text{max}}\, \tau_M$=1. For van der
Waals liquids we have also found that this relation holds to a good
approximation. On the other hand, the two time scales,
$1/(2 \pi \nu_{\text{max}})$ and $\tau_M$, are separated by an order
of magnitude for the monoalcohols displaying a clear signature of the
polymer-like slow mode \cite{Gainaru2014}.  Thus
$2 \pi \nu_{\text{max}}\, \tau_M$ is a decoupling index and deviations
from $2 \pi \nu_{\text{max}}\, \tau_M=1$ quantifies the separation of
the two timescales. The Maxwell relaxation time, $\tau_M$, correspond
to the slowest mode in analysis.  To determine $\tau_M$ one needs to
experimentally reach the low-frequency terminal mode and to reliably
estimate the high frequency plateau of the real part. Since we are
restricted in our frequency window we can not resolve both limiting
regimes for all the temperatures.  In practice, we obtain $\tau_M$
from the fits shown in Fig.\ \ref{fig:dataWithFits}. The model
function has $\tau_M$ as one of the parameters and is able to describe
the low-frequency mode in the data (see Fig.\
\ref{fig:tauMtauAlphaDecoupling}(c)).

In Fig.\ \ref{fig:tauMtauAlphaDecoupling}(a) the separation of the two
relaxation times, defined as $\log \left(2 \pi \nu_{\text{max}}\,
  \tau_M\right)$, is shown for all the investigated concentrations.
For comparison we again show the results of the van der Waals bonded
liquid DC704 and the monoalcohol 2E1H.  The first observation is that
the separation between the two time scales decreases with increasing
water content. This becomes even clearer in Fig.\
\ref{fig:tauMtauAlphaDecoupling}(b) where the separation is shown as a
function of concentration for spectra with their loss peak frequency
close to 100~Hz. There is a sudden drop in time-scale
separation between $x_\text{gly} = 0.6$ and $x_\text{gly} =
0.5$. Another observation from \ref{fig:tauMtauAlphaDecoupling}(a), is
that at concentration above $x_\text{gly} = 0.5$ the time-scale
separation appears to have a temperature dependence, with the
separation increasing as the temperature decreases. In other words the
time-temperature superposition (TTS) does not exactly hold for the
dynamics slower than the alpha process, which can also be seen if one
looks carefully at Fig.\ \ref{fig:shoulder}(c). A zoom af the neat
glycerol data is depicted in Fig.\ \ref{fig:tauMtauAlphaDecoupling}(d)
in order to show the deviation of TTS directly in the spectra.


Regarding the slow mode our results can be summarized as follows: We
demonstrate the presence a slow rheological mode in neat
glycerol. This mode gradually disappears with addition of water. The
separation between the slow mode and the structural relaxation time
scales drops dramatically between $x_\text{gly} = 0.6$ to
$x_\text{gly} = 0.5$. Above $x_\text{gly} = 0.5$ the time scale
separation increases with decreasing temperature. Below $x_\text{gly}
= 0.6$ it is constant or slightly decreasing with decreasing
temperature.

\subsection{High frequency behavior}
\begin{figure}[h!]
  \includegraphics[width=8cm]{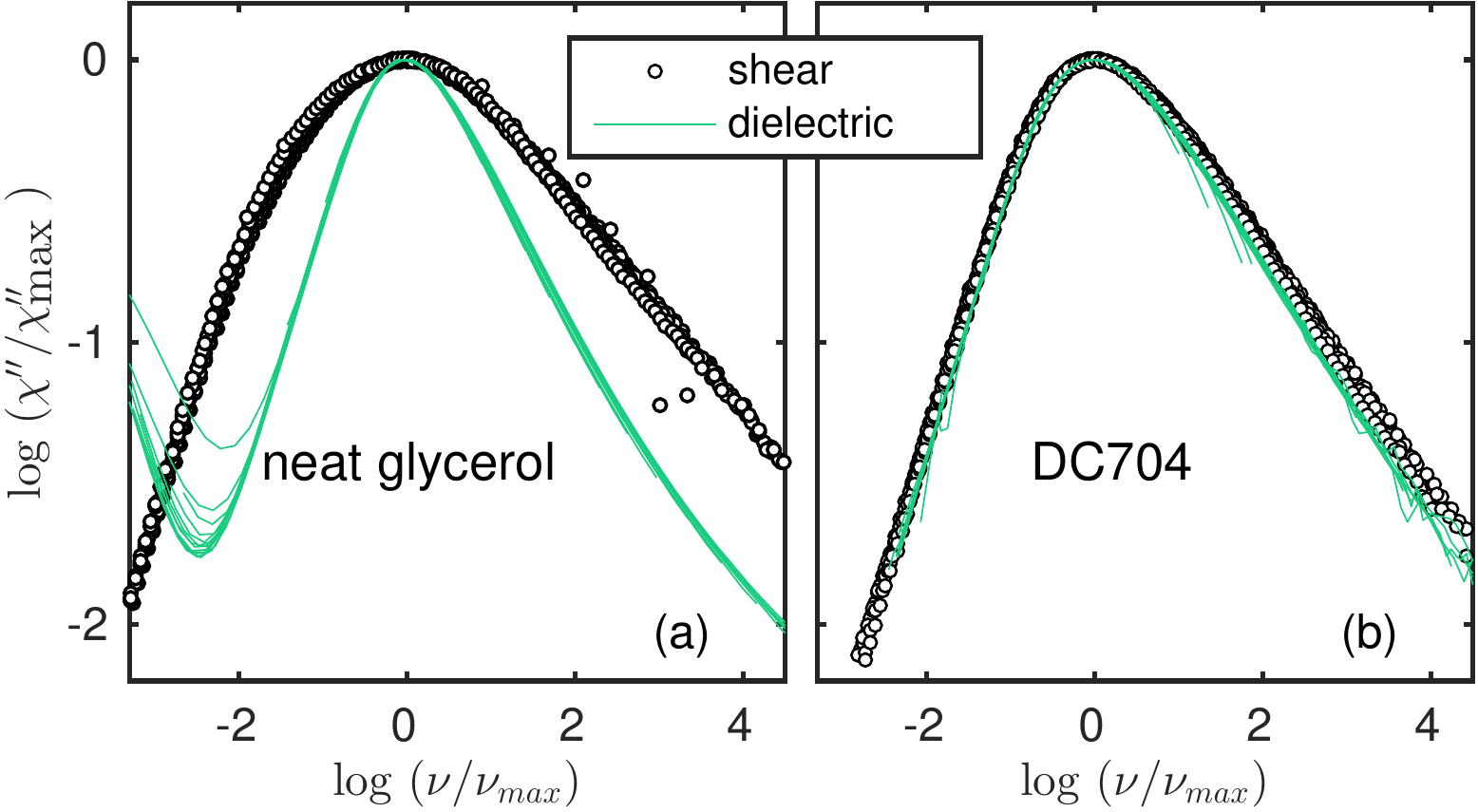}
  \caption{\label{fig:sheardiel}Comparison of shear and dielectric
    spectral shapes for (a) neat glycerol and (b) DC704. For glycerol
    the spectral shapes are very different: the dielectric spectrum is
    narrow and steep while the shear mechanical spectrum is broad due
    to the low frequency shoulder and a less steep high frequency
    slope. In the case of DC704 the observation is the opposite;
    dielectric and shear mechanical spectra for DC704 are nearly
    indistinguishable. This shows that for some liquids different
    probes reveal very different behavior.}
\end{figure}

Another interesting observation, arising when comparing the shear
mechanical spectra to the dielectric ones, relates to the high
frequency flank of the spectrum. The dielectric alpha relaxation peak
of glycerol is often reported to be relatively narrow compared to van
der Waals liquids, as the high frequency flank of the dielectric
spectra is steeper \cite{Bohmer1993}. However, in the shear modulus
loss spectrum the high frequency slope in the double logarithmic plot
is close to $-0.4$ indicating a rather broad underlying
distribution of shear relaxation times. Note that the shear data in
Fig.\ \ref{fig:shoulder}(c) and the dielectric data in Fig.\
\ref{fig:shoulder}(d) have the same scale on the frequency axis. It is
obvious, that the two different response functions are characterized
by different degrees of stretching. This is unlike the case of the van der
Waals liquid DC704 where the dielectric and shear mechanical loss
peaks can barely be distinguished when plottet in this normalized form
\cite{Jakobsen2005}. This striking difference between glycerol and
DC704 is depicted in Fig.\ \ref{fig:sheardiel}.

As observed in Fig.\ \ref{fig:shoulder}(a) a second high frequency
relaxation process becomes clearly visible for $x_\text{gly} = 0.3$
separating from the main relaxation peak with decreasing temperature. This mode 
is absent in the shear mechanical spectra, and thus has a completely
different behavior from high frequency beta-relaxations in neat
systems, which are more pronounced in the shear mechanical spectra
than the dielectric spectra for all systems studied so far
\cite{Jakobsen2005,Niss2005,Jakobsen2011}. 

\section{Discussion}

\subsection{Neat glycerol}

The shear spectra of glycerol resolve an enhanced low frequency
stiffness which is not present in non-associating molecular
liquids. On the other hand, the spectral shape of its corresponding
dielectric susceptibility lacks the complexity of mono-alcohols. Could
it be that in the case of glycerol one encounters slow dynamics which
is rheologically, yet not dielectrically active? A similar behavior
was reported by some of us for 2-ethyl-1-hexylamine
\cite{Adrjanowicz2015}, and to address this question one has to
understand in the first place what kind of structural foundation may
support the emergence of the \emph{dielectric} Debye mode. For
monoalcohols a general agreement has been already established that
they form quasilinear supramolecular structures with the backbone
formed by the hydrogen-bonded hydroxyl groups
\cite{Bohmer2014}. Corroborating the magnitude of the dielectric
strength with this kind of topology, one may rationalize that in
monoalcohols the supramolecular clusters can occur as chains or rings,
depending on steric hindrance of OH-groups. The n-alcohols with the
OH-groups located at the extremity of the molecules are able to
sustain moderately long chains. From the dielectric perspective these
chains can be represented as ``end-to-end'' dipole moments resulting
from the vector summation of the individual dipole moments along the
backbone. The fluctuations of these mesoscopic dipoles are often
considered to be the origin of the Debye mode which resemble to a
certain extent the situation of dielectric normal modes in type A
polymers \cite{Kremer2003}. Within these structural considerations one
may understand why a similar dielectric Debye mode is not generated in
glycerol: with the possibility of several hydrogen bonds per molecule
any preference in the direction of local association vanishes and so
does the effective dipole moment, contrary to the situation for
monoalcohols.

Continuing with the analogy to polymeric behavior it is elucidating to
compare our results to the results of Ref. \cite{Hofman2015} where the
mechanical signature of a linear polymers is compared to that of
dendrimers. While the linear polymers displayed a slow mode clearly
separated from the alpha-relaxation and a distinct rubber plateau in
the real part of the shear modulus, the dendrimers only display
reminiscence of the Rouse dynamics in terms of a low frequency
shoulder on the mechanical loss peak. The difference in the mechanical
spectrum of mono-alcohols and glycerol clearly mimics the behavior
seen for linear polymers and dendrimers depicted in Fig.\ 4 of
Ref. \cite{Hofman2015}. The monoalcohol as the linear polymer exhibits
a clearly separated mode, while in glycerol and dendrimers there is
only a shoulder. This observation supports the interpretation above,
namely that the slow mode and the dielectric Debye-signal in
monoalcohols are due to dynamics associated with a linear structure
formed by hydrogen-bonds, while the slow mode in glycerol is
associated with the branched hydrogen-bonded structures.

\subsection{Glycerol-water mixtures}

The slow polymer-like rheological mode which we observe in glycerol
decreases in significance as water is added and effectively dies out
at equimolar composition. Under the assumption that the slow mode is
due to dynamics of supramolecular hydrogen-bonded structures as argued above,
this tells us that water drastically changes the hydrogen bond structures
and/or their dynamics.

There are several works in literature on the structure of water-glycerol
mixtures, and though most of these studies are carried out at or close
to room temperature we will use these to guide our interpretation of
our findings.

Maybe the most basic structural finding from literature is that the
glycerol-water mixtures have a negative excess molar volume
\citep{Egorov2013}, which means that the total volume decreases upon
mixing.  The excess molar volume has a minimum close to
$x_\text{gly} \approx 0.3$ coinciding with the minimum in the melting
transition temperature (Fig. \ref{fig:TgvsXg}). The negative excess
molar volume indicates that there is an increased packing efficiency
as the water concentration increases in the range studied in this
work. Moreover, it has been shown that the internal structure of
glycerol itself changes when mixed with water \citep{Egorov2011},
which might also have a direct effect on the molecular packing
efficiency.

Towey et.\ al.\ used neutron scattering in combination with Reverse
Monte Carlo (RMC) simulations to study the hydrogen-bonded structure in
glycerol \citep{Towey2011} as well as in glycerol-water mixtures
\citep{Towey2011a,Towey2012,Towey2012b,Towey2013}. They find that in
neat glycerol each molecule forms on average $5.68\pm 1.51$ hydrogen
bonds, while for neat water the average number of hydrogen bonds per
molecule is about $3.56\pm 1.10$. Mixing water and glycerol preserves
to a large extent the number of hydrogen bonds per molecule although
the number of glycerol-glycerol hydrogen bonds decreases progressively
as they are replaced by glycerol-water bonds, so the hydrogen-bonded
structure changes.

We assume that the changes we see in the mechanical spectra among the
different compositions must be due to changes in the dynamics of the
concentration dependent hydrogen-bonded network. The sudden change at
glycerol concentration $0.5<X_\text{gly} < 0.6$ suggests a qualitative
change of the network behavior around this concentration.  Indeed, in
Ref. \cite{Towey2013} Towey et al.\ studied the cluster size
distribution and revealed that percolating water clusters are found at
mole fractions $x_\text{gly}=0.5$ and below, while percolating
glycerol clusters are found from $x_\text{gly}=0.25$ and above. As
mentioned above, one should be careful when comparing our findings to
the results of Towey et al.\ which are obtained with the aid of RMC
and data taken at room temperature, while we study deeply supercooled
samples. Nevertheless, it is striking that the concentration
$x_\text{gly}\simeq 0.5$ where we observe a dynamical change coincides
with a structural change for the water clusters in terms of
percolation, while nothing dramatic happens to the glycerol-glycerol
network. An interpretation could be that water acts as a
``lubricant'', reducing the low frequency stiffness of the glycerol
network.

In Ref. \cite{Hayashi2006} it is noted that the dielectric spectra
obey time frequency superposition for glycerol concentrations above
0.55 while this superposition breaks down at $x_\text{gly} = 0.55$,
and a clear high frequency mode appears when the glycerol
concentration is below $x_\text{gly} = 0.4$.  It is tempting to
ascribe this fast dynamics to the Johari-Goldstein beta-process which
is generally exhibited by neat glass-forming systems near their glassy
state. However, this high frequency spectral contribution is absent in
the complementary shear mechanical spectra. Since the beta-relaxation
generally is more pronounced in the mechanical spectra than in the
dielectric ones \cite{Jakobsen2005, Niss2005, Jakobsen2011}, it
becomes clear that this water-induced relaxation mode cannot be
regarded as a ubiquitous secondary process. It has also earlier been
suggested that the high frequency dielectric shoulder is a signature
of water-water dynamics in aggregated water domains
\cite{Hayashi2006}.

In water-alcohol mixtures in general hydrogen bonding is known as a
driving force for clustering in the liquid and responsible of the
limit of miscibility \cite{Artola2013}; compared to these studies in
the low viscosity (high T) regime, the effect of decreasing
temperature will be to enhance the phase separation. As the water
concentration is increased we expect that there are increasing
concentration fluctuations, and nano-segregation is likely to occur
close to $T_g$ in the mixtures with high water concentrations. Further
structural investigations need to be considered for the elucidation of
this peculiar mixing behavior and its connection with other anomalies
of water-related materials.

\section{Conclusion}
In this work, we demonstrate the existence of a slow rheological mode
in neat glycerol. The slow relaxation feature is similar to the
suprasegmental dynamics seen in dendrimers and we attribute it to the
rearrangements of the branced hydrogen-bonded network. To our
knowledge this is the first demonstration of hydrogen-bonded network
dynamics in a poly-alcohol, while similar signals have been reported
so far in monoalcohols only.  The slow mode becomes weaker when adding
water, and disappears quite abruptly at equimolar composition. This
corresponds to the water concentration where percolating water
clusters appear according to RMC simulations of neutron scattering
data. We speculate that water acts as a lubricant softening the
modes of the glycerol-glycerol hydrogen-bonded matrix. Our results
underlines how different parts of the dynamics in hydrogen-bonded
systems are evident from different types of spectroscopy making the
use of complementary techniques imperative in this field.

\section{Acknowledgements}
We thank Ivan Popov and Prof. Yuri Feldmann for making the dielectric
data of Ref.\ \cite{Popov2015} available in an electronic form. The
following funding bodies are acknowledged: Danish Council for
Independent Research (Sapere Aude: Starting Grant), the Danish
National Research Foundation (DNRF) and Laboratoire Léon Brillouin
(UMR CEA-CNRS, Paris-Saclay).

%
\bibliography{GlyWat}

\begin{thebibliography}{64}
\expandafter\ifx\csname natexlab\endcsname\relax\def\natexlab#1{#1}\fi
\expandafter\ifx\csname bibnamefont\endcsname\relax
  \def\bibnamefont#1{#1}\fi
\expandafter\ifx\csname bibfnamefont\endcsname\relax
  \def\bibfnamefont#1{#1}\fi
\expandafter\ifx\csname citenamefont\endcsname\relax
  \def\citenamefont#1{#1}\fi
\expandafter\ifx\csname url\endcsname\relax
  \def\url#1{\texttt{#1}}\fi
\expandafter\ifx\csname urlprefix\endcsname\relax\def\urlprefix{URL }\fi
\providecommand{\bibinfo}[2]{#2}
\providecommand{\eprint}[2][]{\url{#2}}

\bibitem[{\citenamefont{Pauling}(1939)}]{Pauling1939}
\bibinfo{author}{\bibfnamefont{L.}~\bibnamefont{Pauling}},
  \emph{\bibinfo{title}{The Nature of the Chemical Bond and The Structure of
  Molecules and Crystals}} (\bibinfo{publisher}{Cornell University Press,
  Ithaca, NY}, \bibinfo{year}{1939}).

\bibitem[{\citenamefont{Jeffrey and Saenger}(1991)}]{Jeffrey1991}
\bibinfo{author}{\bibfnamefont{G.~A.} \bibnamefont{Jeffrey}} \bibnamefont{and}
  \bibinfo{author}{\bibfnamefont{W.}~\bibnamefont{Saenger}},
  \emph{\bibinfo{title}{Hydrogen Bonding in Biological Structures}}
  (\bibinfo{publisher}{Springer-Verlag, Berlin}, \bibinfo{year}{1991}).

\bibitem[{\citenamefont{Desiraju and Steiner}(1999)}]{Desiraju1999}
\bibinfo{author}{\bibfnamefont{G.~R.} \bibnamefont{Desiraju}} \bibnamefont{and}
  \bibinfo{author}{\bibfnamefont{T.}~\bibnamefont{Steiner}},
  \emph{\bibinfo{title}{The Weak Hydrogen Bond in Structural Chemistry and
  Biology}} (\bibinfo{publisher}{Oxford University Press, Chichester},
  \bibinfo{year}{1999}).

\bibitem[{\citenamefont{Mar\'echal}(2007)}]{Waterbook2007}
\bibinfo{author}{\bibfnamefont{Y.}~\bibnamefont{Mar\'echal}},
  \emph{\bibinfo{title}{The hydrogen bond and the water molecule: The physics
  and chemistry of water, aqueous and bio-media}}
  (\bibinfo{publisher}{Elsevier}, \bibinfo{address}{Amsterdam},
  \bibinfo{year}{2007}).

\bibitem[{\citenamefont{Gibson and Giauque}(1923)}]{Gibson1923}
\bibinfo{author}{\bibfnamefont{G.~E.} \bibnamefont{Gibson}} \bibnamefont{and}
  \bibinfo{author}{\bibfnamefont{W.~F.} \bibnamefont{Giauque}},
  \textbf{\bibinfo{volume}{45}}, \bibinfo{pages}{93} (\bibinfo{year}{1923}),
  ISSN \bibinfo{issn}{0002-7863}.

\bibitem[{\citenamefont{Wang and Wright}(1971)}]{Wang1971}
\bibinfo{author}{\bibfnamefont{C.~H.} \bibnamefont{Wang}} \bibnamefont{and}
  \bibinfo{author}{\bibfnamefont{R.~B.} \bibnamefont{Wright}},
  \bibinfo{journal}{J. Chem. Phys.} \textbf{\bibinfo{volume}{55}},
  \bibinfo{pages}{1617} (\bibinfo{year}{1971}).

\bibitem[{\citenamefont{Schiener et~al.}(1996)\citenamefont{Schiener,
  B\"{o}hmer, Loidl, and Chamberlin}}]{Schiener1996}
\bibinfo{author}{\bibfnamefont{B.}~\bibnamefont{Schiener}},
  \bibinfo{author}{\bibfnamefont{R.}~\bibnamefont{B\"{o}hmer}},
  \bibinfo{author}{\bibfnamefont{A.}~\bibnamefont{Loidl}}, \bibnamefont{and}
  \bibinfo{author}{\bibfnamefont{R.}~\bibnamefont{Chamberlin}},
  \bibinfo{journal}{Science} \textbf{\bibinfo{volume}{274}},
  \bibinfo{pages}{752} (\bibinfo{year}{1996}), ISSN \bibinfo{issn}{0036-8075}.

\bibitem[{\citenamefont{Berthier et~al.}(2005)\citenamefont{Berthier, Biroli,
  Bouchaud, Cipelletti, Masri, L{\textquoteright}H{\^o}te, Ladieu, and
  Pierno}}]{Berthier2005}
\bibinfo{author}{\bibfnamefont{L.}~\bibnamefont{Berthier}},
  \bibinfo{author}{\bibfnamefont{G.}~\bibnamefont{Biroli}},
  \bibinfo{author}{\bibfnamefont{J.-P.} \bibnamefont{Bouchaud}},
  \bibinfo{author}{\bibfnamefont{L.}~\bibnamefont{Cipelletti}},
  \bibinfo{author}{\bibfnamefont{D.~E.} \bibnamefont{Masri}},
  \bibinfo{author}{\bibfnamefont{D.}~\bibnamefont{L{\textquoteright}H{\^o}te}},
  \bibinfo{author}{\bibfnamefont{F.}~\bibnamefont{Ladieu}}, \bibnamefont{and}
  \bibinfo{author}{\bibfnamefont{M.}~\bibnamefont{Pierno}},
  \bibinfo{journal}{Science} \textbf{\bibinfo{volume}{310}},
  \bibinfo{pages}{1797} (\bibinfo{year}{2005}), ISSN \bibinfo{issn}{0036-8075}.

\bibitem[{\citenamefont{Pezeril et~al.}(2009)\citenamefont{Pezeril, Klieber,
  Andrieu, and Nelson}}]{Pezeril2009}
\bibinfo{author}{\bibfnamefont{T.}~\bibnamefont{Pezeril}},
  \bibinfo{author}{\bibfnamefont{C.}~\bibnamefont{Klieber}},
  \bibinfo{author}{\bibfnamefont{S.}~\bibnamefont{Andrieu}}, \bibnamefont{and}
  \bibinfo{author}{\bibfnamefont{K.~A.} \bibnamefont{Nelson}},
  \bibinfo{journal}{Phys. Rev. Lett.} \textbf{\bibinfo{volume}{102}},
  \bibinfo{pages}{107402} (\bibinfo{year}{2009}).

\bibitem[{\citenamefont{Albert et~al.}(2016)\citenamefont{Albert, Bauer, Michl,
  Biroli, Bouchaud, Loidl, Lunkenheimer, Tourbot, Wiertel-Gasquet, and
  Ladieu}}]{Albert2016}
\bibinfo{author}{\bibfnamefont{S.}~\bibnamefont{Albert}},
  \bibinfo{author}{\bibfnamefont{T.}~\bibnamefont{Bauer}},
  \bibinfo{author}{\bibfnamefont{M.}~\bibnamefont{Michl}},
  \bibinfo{author}{\bibfnamefont{G.}~\bibnamefont{Biroli}},
  \bibinfo{author}{\bibfnamefont{J.-P.} \bibnamefont{Bouchaud}},
  \bibinfo{author}{\bibfnamefont{A.}~\bibnamefont{Loidl}},
  \bibinfo{author}{\bibfnamefont{P.}~\bibnamefont{Lunkenheimer}},
  \bibinfo{author}{\bibfnamefont{R.}~\bibnamefont{Tourbot}},
  \bibinfo{author}{\bibfnamefont{C.}~\bibnamefont{Wiertel-Gasquet}},
  \bibnamefont{and} \bibinfo{author}{\bibfnamefont{F.}~\bibnamefont{Ladieu}},
  \bibinfo{journal}{Science} \textbf{\bibinfo{volume}{352}},
  \bibinfo{pages}{1308} (\bibinfo{year}{2016}), ISSN \bibinfo{issn}{0036-8075}.

\bibitem[{\citenamefont{Salt}(1958)}]{Salt1958}
\bibinfo{author}{\bibfnamefont{R.~W.} \bibnamefont{Salt}},
  \bibinfo{journal}{Nature} \textbf{\bibinfo{volume}{181}},
  \bibinfo{pages}{1281} (\bibinfo{year}{1958}).

\bibitem[{\citenamefont{Dashnau et~al.}(2006)\citenamefont{Dashnau, Nucci,
  Sharp, and Vanderkooi}}]{Dashnau2006}
\bibinfo{author}{\bibfnamefont{J.~L.} \bibnamefont{Dashnau}},
  \bibinfo{author}{\bibfnamefont{N.~V.} \bibnamefont{Nucci}},
  \bibinfo{author}{\bibfnamefont{K.~A.} \bibnamefont{Sharp}}, \bibnamefont{and}
  \bibinfo{author}{\bibfnamefont{J.~M.} \bibnamefont{Vanderkooi}},
  \bibinfo{journal}{J. Phys. Chem. B} \textbf{\bibinfo{volume}{110}},
  \bibinfo{pages}{13670} (\bibinfo{year}{2006}).

\bibitem[{\citenamefont{Li et~al.}(2008)\citenamefont{Li, Liu, shu Liu, and
  lung Chen}}]{Li2008}
\bibinfo{author}{\bibfnamefont{D.-X.} \bibnamefont{Li}},
  \bibinfo{author}{\bibfnamefont{B.-L.} \bibnamefont{Liu}},
  \bibinfo{author}{\bibfnamefont{Y.}~\bibnamefont{shu Liu}}, \bibnamefont{and}
  \bibinfo{author}{\bibfnamefont{C.}~\bibnamefont{lung Chen}},
  \bibinfo{journal}{Cryobiology} \textbf{\bibinfo{volume}{56}},
  \bibinfo{pages}{114 } (\bibinfo{year}{2008}), ISSN \bibinfo{issn}{0011-2240}.

\bibitem[{\citenamefont{Popov et~al.}(2015)\citenamefont{Popov, Gutina,
  Sokolov, and Feldman}}]{Popov2015}
\bibinfo{author}{\bibfnamefont{I.}~\bibnamefont{Popov}},
  \bibinfo{author}{\bibfnamefont{A.~G.} \bibnamefont{Gutina}},
  \bibinfo{author}{\bibfnamefont{A.~P.} \bibnamefont{Sokolov}},
  \bibnamefont{and} \bibinfo{author}{\bibfnamefont{Y.}~\bibnamefont{Feldman}},
  \bibinfo{journal}{Phys. Chem. Chem. Phys.} \textbf{\bibinfo{volume}{17}},
  \bibinfo{pages}{1} (\bibinfo{year}{2015}).

\bibitem[{\citenamefont{Murata and Tanaka}(2012)}]{Murata2012}
\bibinfo{author}{\bibfnamefont{K.-i.} \bibnamefont{Murata}} \bibnamefont{and}
  \bibinfo{author}{\bibfnamefont{H.}~\bibnamefont{Tanaka}},
  \bibinfo{journal}{Nat. Mater.} \textbf{\bibinfo{volume}{11}},
  \bibinfo{pages}{436} (\bibinfo{year}{2012}).

\bibitem[{\citenamefont{Lunkenheimer et~al.}(2000)\citenamefont{Lunkenheimer,
  Schneider, Brand, and Loid}}]{Lunkenheimer2000}
\bibinfo{author}{\bibfnamefont{P.}~\bibnamefont{Lunkenheimer}},
  \bibinfo{author}{\bibfnamefont{U.}~\bibnamefont{Schneider}},
  \bibinfo{author}{\bibfnamefont{R.}~\bibnamefont{Brand}}, \bibnamefont{and}
  \bibinfo{author}{\bibfnamefont{A.}~\bibnamefont{Loid}},
  \bibinfo{journal}{Contemp. Phys.} \textbf{\bibinfo{volume}{41}},
  \bibinfo{pages}{15} (\bibinfo{year}{2000}).

\bibitem[{\citenamefont{Gainaru et~al.}(2005)\citenamefont{Gainaru, Rivera,
  Putselyk, Eska, and R\"ossler}}]{Gainaru2005}
\bibinfo{author}{\bibfnamefont{C.}~\bibnamefont{Gainaru}},
  \bibinfo{author}{\bibfnamefont{A.}~\bibnamefont{Rivera}},
  \bibinfo{author}{\bibfnamefont{S.}~\bibnamefont{Putselyk}},
  \bibinfo{author}{\bibfnamefont{G.}~\bibnamefont{Eska}}, \bibnamefont{and}
  \bibinfo{author}{\bibfnamefont{E.~A.} \bibnamefont{R\"ossler}},
  \bibinfo{journal}{Phys. Rev. B} \textbf{\bibinfo{volume}{72}},
  \bibinfo{pages}{174203} (\bibinfo{year}{2005}).

\bibitem[{\citenamefont{Sudo et~al.}(2002)\citenamefont{Sudo, Shimomura,
  Shinyashiki, and Yagihara}}]{Sudo2002}
\bibinfo{author}{\bibfnamefont{S.}~\bibnamefont{Sudo}},
  \bibinfo{author}{\bibfnamefont{M.}~\bibnamefont{Shimomura}},
  \bibinfo{author}{\bibfnamefont{N.}~\bibnamefont{Shinyashiki}},
  \bibnamefont{and} \bibinfo{author}{\bibfnamefont{S.}~\bibnamefont{Yagihara}},
  \bibinfo{journal}{J. Non-Cryst. Solids} \textbf{\bibinfo{volume}{307–310}},
  \bibinfo{pages}{356 } (\bibinfo{year}{2002}), ISSN \bibinfo{issn}{0022-3093}.

\bibitem[{\citenamefont{Puzenko et~al.}(2005)\citenamefont{Puzenko, Hayashi,
  Ryabov, Balin, Feldman, Kaatze, and Behrends}}]{Puzenko2005}
\bibinfo{author}{\bibfnamefont{A.}~\bibnamefont{Puzenko}},
  \bibinfo{author}{\bibfnamefont{Y.}~\bibnamefont{Hayashi}},
  \bibinfo{author}{\bibfnamefont{Y.~E.} \bibnamefont{Ryabov}},
  \bibinfo{author}{\bibfnamefont{I.}~\bibnamefont{Balin}},
  \bibinfo{author}{\bibfnamefont{Y.}~\bibnamefont{Feldman}},
  \bibinfo{author}{\bibfnamefont{U.}~\bibnamefont{Kaatze}}, \bibnamefont{and}
  \bibinfo{author}{\bibfnamefont{R.}~\bibnamefont{Behrends}},
  \bibinfo{journal}{J. Phys. Chem. B} \textbf{\bibinfo{volume}{109}},
  \bibinfo{pages}{6031} (\bibinfo{year}{2005}), \bibinfo{note}{pMID: 16851659}.

\bibitem[{\citenamefont{Puzenko et~al.}(2007)\citenamefont{Puzenko, Hayashi,
  and Feldman}}]{Puzenko2007}
\bibinfo{author}{\bibfnamefont{A.}~\bibnamefont{Puzenko}},
  \bibinfo{author}{\bibfnamefont{Y.}~\bibnamefont{Hayashi}}, \bibnamefont{and}
  \bibinfo{author}{\bibfnamefont{Y.}~\bibnamefont{Feldman}},
  \bibinfo{journal}{J. Non-Cryst. Solids} \textbf{\bibinfo{volume}{353}},
  \bibinfo{pages}{4518 } (\bibinfo{year}{2007}), ISSN
  \bibinfo{issn}{0022-3093}.

\bibitem[{\citenamefont{Hayashi et~al.}(2005)\citenamefont{Hayashi, Puzenko,
  Balin, Ryabov, and Feldman}}]{Hayashi2005}
\bibinfo{author}{\bibfnamefont{Y.}~\bibnamefont{Hayashi}},
  \bibinfo{author}{\bibfnamefont{A.}~\bibnamefont{Puzenko}},
  \bibinfo{author}{\bibfnamefont{I.}~\bibnamefont{Balin}},
  \bibinfo{author}{\bibfnamefont{Y.~E.} \bibnamefont{Ryabov}},
  \bibnamefont{and} \bibinfo{author}{\bibfnamefont{Y.}~\bibnamefont{Feldman}},
  \bibinfo{journal}{J. Phys. Chem. B} \textbf{\bibinfo{volume}{109}},
  \bibinfo{pages}{9174} (\bibinfo{year}{2005}).

\bibitem[{\citenamefont{Hayashi et~al.}(2006)\citenamefont{Hayashi, Puzenko,
  and Feldman}}]{Hayashi2006}
\bibinfo{author}{\bibfnamefont{Y.}~\bibnamefont{Hayashi}},
  \bibinfo{author}{\bibfnamefont{A.}~\bibnamefont{Puzenko}}, \bibnamefont{and}
  \bibinfo{author}{\bibfnamefont{Y.}~\bibnamefont{Feldman}},
  \bibinfo{journal}{J. Non-Cryst. Solids} \textbf{\bibinfo{volume}{352}},
  \bibinfo{pages}{4696 } (\bibinfo{year}{2006}), ISSN
  \bibinfo{issn}{0022-3093}, \bibinfo{note}{proceedings of the 5th
  International Discussion Meeting on Relaxations in Complex Systems5th
  International Discussion Meeting on Relaxations in Complex Systems}.

\bibitem[{\citenamefont{Murthy and Nayak}(1993)}]{Murthy1993}
\bibinfo{author}{\bibfnamefont{S.~S.~N.} \bibnamefont{Murthy}}
  \bibnamefont{and} \bibinfo{author}{\bibfnamefont{S.~K.} \bibnamefont{Nayak}},
  \bibinfo{journal}{J. Chem. Phys.} \textbf{\bibinfo{volume}{99}},
  \bibinfo{pages}{5362} (\bibinfo{year}{1993}).

\bibitem[{\citenamefont{Huth et~al.}(2007)\citenamefont{Huth, Wang, Schick, and
  Richert}}]{Huth2007}
\bibinfo{author}{\bibfnamefont{H.}~\bibnamefont{Huth}},
  \bibinfo{author}{\bibfnamefont{L.-M.} \bibnamefont{Wang}},
  \bibinfo{author}{\bibfnamefont{C.}~\bibnamefont{Schick}}, \bibnamefont{and}
  \bibinfo{author}{\bibfnamefont{R.}~\bibnamefont{Richert}},
  \bibinfo{journal}{J. Chem. Phys.} \textbf{\bibinfo{volume}{126}},
  \bibinfo{pages}{104503} (\bibinfo{year}{2007}).

\bibitem[{\citenamefont{B\"{o}hmer et~al.}(2014)\citenamefont{B\"{o}hmer,
  Gainaru, and Richert}}]{Bohmer2014}
\bibinfo{author}{\bibfnamefont{R.}~\bibnamefont{B\"{o}hmer}},
  \bibinfo{author}{\bibfnamefont{C.}~\bibnamefont{Gainaru}}, \bibnamefont{and}
  \bibinfo{author}{\bibfnamefont{R.}~\bibnamefont{Richert}},
  \bibinfo{journal}{Phys. Rep.} \textbf{\bibinfo{volume}{545}},
  \bibinfo{pages}{125} (\bibinfo{year}{2014}).

\bibitem[{\citenamefont{Singh et~al.}(2013)\citenamefont{Singh,
  Alba-Simionesco, and Richert}}]{Singh2013}
\bibinfo{author}{\bibfnamefont{L.~P.} \bibnamefont{Singh}},
  \bibinfo{author}{\bibfnamefont{C.}~\bibnamefont{Alba-Simionesco}},
  \bibnamefont{and} \bibinfo{author}{\bibfnamefont{R.}~\bibnamefont{Richert}},
  \bibinfo{journal}{J. Chem. Phys.} \textbf{\bibinfo{volume}{139}},
  \bibinfo{pages}{144503} (\bibinfo{year}{2013}).

\bibitem[{\citenamefont{Singh et~al.}(2015)\citenamefont{Singh, Raihane,
  Alba-Simionesco, and Richert}}]{Singh2015}
\bibinfo{author}{\bibfnamefont{L.~P.} \bibnamefont{Singh}},
  \bibinfo{author}{\bibfnamefont{A.}~\bibnamefont{Raihane}},
  \bibinfo{author}{\bibfnamefont{C.}~\bibnamefont{Alba-Simionesco}},
  \bibnamefont{and} \bibinfo{author}{\bibfnamefont{R.}~\bibnamefont{Richert}},
  \bibinfo{journal}{J. Chem. Phys.} \textbf{\bibinfo{volume}{142}},
  \bibinfo{pages}{014501} (\bibinfo{year}{2015}).

\bibitem[{\citenamefont{Morineau and Alba-Simionesco}(1998)}]{Morineau1998}
\bibinfo{author}{\bibfnamefont{D.}~\bibnamefont{Morineau}} \bibnamefont{and}
  \bibinfo{author}{\bibfnamefont{C.}~\bibnamefont{Alba-Simionesco}},
  \bibinfo{journal}{J. Chem. Phys.} \textbf{\bibinfo{volume}{109}},
  \bibinfo{pages}{8494} (\bibinfo{year}{1998}).

\bibitem[{\citenamefont{Mandanici et~al.}(2005)\citenamefont{Mandanici,
  Cutroni, and Richert}}]{Mandanici2005}
\bibinfo{author}{\bibfnamefont{A.}~\bibnamefont{Mandanici}},
  \bibinfo{author}{\bibfnamefont{M.}~\bibnamefont{Cutroni}}, \bibnamefont{and}
  \bibinfo{author}{\bibfnamefont{R.}~\bibnamefont{Richert}},
  \bibinfo{journal}{J. Chem. Phys.} \textbf{\bibinfo{volume}{122}},
  \bibinfo{pages}{084508} (\bibinfo{year}{2005}).

\bibitem[{\citenamefont{Gao et~al.}(2013)\citenamefont{Gao, Bi, Li, Liu, Tian,
  and Wang}}]{Gao2013}
\bibinfo{author}{\bibfnamefont{Y.}~\bibnamefont{Gao}},
  \bibinfo{author}{\bibfnamefont{D.}~\bibnamefont{Bi}},
  \bibinfo{author}{\bibfnamefont{X.}~\bibnamefont{Li}},
  \bibinfo{author}{\bibfnamefont{R.}~\bibnamefont{Liu}},
  \bibinfo{author}{\bibfnamefont{Y.}~\bibnamefont{Tian}}, \bibnamefont{and}
  \bibinfo{author}{\bibfnamefont{L.-M.} \bibnamefont{Wang}},
  \bibinfo{journal}{J. Chem. Phys.} \textbf{\bibinfo{volume}{139}},
  \bibinfo{pages}{024503} (\bibinfo{year}{2013}).

\bibitem[{\citenamefont{Wang and Richert}(2005)}]{Wang2005}
\bibinfo{author}{\bibfnamefont{L.-M.} \bibnamefont{Wang}} \bibnamefont{and}
  \bibinfo{author}{\bibfnamefont{R.}~\bibnamefont{Richert}},
  \bibinfo{journal}{J. Chem. Phys.} \textbf{\bibinfo{volume}{123}},
  \bibinfo{pages}{054516} (\bibinfo{year}{2005}).

\bibitem[{\citenamefont{Cosby et~al.}(2015)\citenamefont{Cosby, Holt, Griffin,
  Wang, and Sangoro}}]{Cosby2015}
\bibinfo{author}{\bibfnamefont{T.}~\bibnamefont{Cosby}},
  \bibinfo{author}{\bibfnamefont{A.}~\bibnamefont{Holt}},
  \bibinfo{author}{\bibfnamefont{P.~J.} \bibnamefont{Griffin}},
  \bibinfo{author}{\bibfnamefont{Y.}~\bibnamefont{Wang}}, \bibnamefont{and}
  \bibinfo{author}{\bibfnamefont{J.}~\bibnamefont{Sangoro}},
  \bibinfo{journal}{J. Phys. Chem. Lett.} \textbf{\bibinfo{volume}{6}},
  \bibinfo{pages}{3961} (\bibinfo{year}{2015}), \bibinfo{note}{pMID: 26722899}.

\bibitem[{\citenamefont{Adrjanowicz et~al.}(2013)\citenamefont{Adrjanowicz,
  Kaminski, Dulski, Wlodarczyk, Bartkowiak, Popenda, Jurga, Kujawski, Kruk,
  Bernard et~al.}}]{Adrjanowicz2013}
\bibinfo{author}{\bibfnamefont{K.}~\bibnamefont{Adrjanowicz}},
  \bibinfo{author}{\bibfnamefont{K.}~\bibnamefont{Kaminski}},
  \bibinfo{author}{\bibfnamefont{M.}~\bibnamefont{Dulski}},
  \bibinfo{author}{\bibfnamefont{P.}~\bibnamefont{Wlodarczyk}},
  \bibinfo{author}{\bibfnamefont{G.}~\bibnamefont{Bartkowiak}},
  \bibinfo{author}{\bibfnamefont{L.}~\bibnamefont{Popenda}},
  \bibinfo{author}{\bibfnamefont{S.}~\bibnamefont{Jurga}},
  \bibinfo{author}{\bibfnamefont{J.}~\bibnamefont{Kujawski}},
  \bibinfo{author}{\bibfnamefont{J.}~\bibnamefont{Kruk}},
  \bibinfo{author}{\bibfnamefont{M.~K.} \bibnamefont{Bernard}},
  \bibnamefont{et~al.}, \bibinfo{journal}{J. Chem. Phys.}
  \textbf{\bibinfo{volume}{139}}, \bibinfo{pages}{111103}
  (\bibinfo{year}{2013}).

\bibitem[{\citenamefont{Rams-Baron et~al.}(2015)\citenamefont{Rams-Baron,
  Wojnarowska, Dulski, Ratuszna, and Paluch}}]{Rams-Baron2015}
\bibinfo{author}{\bibfnamefont{M.}~\bibnamefont{Rams-Baron}},
  \bibinfo{author}{\bibfnamefont{Z.}~\bibnamefont{Wojnarowska}},
  \bibinfo{author}{\bibfnamefont{M.}~\bibnamefont{Dulski}},
  \bibinfo{author}{\bibfnamefont{A.}~\bibnamefont{Ratuszna}}, \bibnamefont{and}
  \bibinfo{author}{\bibfnamefont{M.}~\bibnamefont{Paluch}},
  \bibinfo{journal}{Phys. Rev. E} \textbf{\bibinfo{volume}{92}},
  \bibinfo{pages}{022309} (\bibinfo{year}{2015}).

\bibitem[{\citenamefont{Pawlus et~al.}(2013)\citenamefont{Pawlus, Klotz, and
  Paluch}}]{Pawlus2013}
\bibinfo{author}{\bibfnamefont{S.}~\bibnamefont{Pawlus}},
  \bibinfo{author}{\bibfnamefont{S.}~\bibnamefont{Klotz}}, \bibnamefont{and}
  \bibinfo{author}{\bibfnamefont{M.}~\bibnamefont{Paluch}},
  \bibinfo{journal}{Phys. Rev. Lett.} \textbf{\bibinfo{volume}{110}},
  \bibinfo{pages}{173004} (\bibinfo{year}{2013}).

\bibitem[{\citenamefont{Gainaru et~al.}(2014)\citenamefont{Gainaru, Figuli,
  Hecksher, Jakobsen, Dyre, Wilhelm, and B\"ohmer}}]{Gainaru2014}
\bibinfo{author}{\bibfnamefont{C.}~\bibnamefont{Gainaru}},
  \bibinfo{author}{\bibfnamefont{R.}~\bibnamefont{Figuli}},
  \bibinfo{author}{\bibfnamefont{T.}~\bibnamefont{Hecksher}},
  \bibinfo{author}{\bibfnamefont{B.}~\bibnamefont{Jakobsen}},
  \bibinfo{author}{\bibfnamefont{J.~C.} \bibnamefont{Dyre}},
  \bibinfo{author}{\bibfnamefont{M.}~\bibnamefont{Wilhelm}}, \bibnamefont{and}
  \bibinfo{author}{\bibfnamefont{R.}~\bibnamefont{B\"ohmer}},
  \bibinfo{journal}{Phys. Rev. Lett.} \textbf{\bibinfo{volume}{112}},
  \bibinfo{pages}{098301} (\bibinfo{year}{2014}).

\bibitem[{\citenamefont{Hecksher}(2016)}]{Hecksher2016}
\bibinfo{author}{\bibfnamefont{T.}~\bibnamefont{Hecksher}},
  \bibinfo{journal}{J. Chem. Phys.} \textbf{\bibinfo{volume}{144}}
  (\bibinfo{year}{2016}).

\bibitem[{\citenamefont{Ferry}(1980)}]{Ferry1980}
\bibinfo{author}{\bibfnamefont{J.~D.} \bibnamefont{Ferry}},
  \emph{\bibinfo{title}{Viscoelastic Properties of Polymers}}
  (\bibinfo{publisher}{Wiley}, \bibinfo{address}{New York},
  \bibinfo{year}{1980}).

\bibitem[{\citenamefont{Adrjanowicz et~al.}(2015)\citenamefont{Adrjanowicz,
  Jakobsen, Hecksher, Kaminski, Dulski, Paluch, and Niss}}]{Adrjanowicz2015}
\bibinfo{author}{\bibfnamefont{K.}~\bibnamefont{Adrjanowicz}},
  \bibinfo{author}{\bibfnamefont{B.}~\bibnamefont{Jakobsen}},
  \bibinfo{author}{\bibfnamefont{T.}~\bibnamefont{Hecksher}},
  \bibinfo{author}{\bibfnamefont{K.}~\bibnamefont{Kaminski}},
  \bibinfo{author}{\bibfnamefont{M.}~\bibnamefont{Dulski}},
  \bibinfo{author}{\bibfnamefont{M.}~\bibnamefont{Paluch}}, \bibnamefont{and}
  \bibinfo{author}{\bibfnamefont{K.}~\bibnamefont{Niss}}, \bibinfo{journal}{J.
  Chem. Phys.} \textbf{\bibinfo{volume}{143}} (\bibinfo{year}{2015}).

\bibitem[{\citenamefont{Hecksher and Jakobsen}(2014)}]{Hecksher2014}
\bibinfo{author}{\bibfnamefont{T.}~\bibnamefont{Hecksher}} \bibnamefont{and}
  \bibinfo{author}{\bibfnamefont{B.}~\bibnamefont{Jakobsen}},
  \bibinfo{journal}{J. Chem. Phys.} \textbf{\bibinfo{volume}{141}},
  \bibinfo{pages}{101104} (\bibinfo{year}{2014}).

\bibitem[{\citenamefont{Christensen and Olsen}(1995)}]{Christensen1995}
\bibinfo{author}{\bibfnamefont{T.}~\bibnamefont{Christensen}} \bibnamefont{and}
  \bibinfo{author}{\bibfnamefont{N.~B.} \bibnamefont{Olsen}},
  \bibinfo{journal}{Rev. Sci. Instrum.} \textbf{\bibinfo{volume}{66}},
  \bibinfo{pages}{5019} (\bibinfo{year}{1995}).

\bibitem[{\citenamefont{Hecksher et~al.}(2017)\citenamefont{Hecksher, Olsen,
  and Dyre}}]{Hecksher2017}
\bibinfo{author}{\bibfnamefont{T.}~\bibnamefont{Hecksher}},
  \bibinfo{author}{\bibfnamefont{N.~B.} \bibnamefont{Olsen}}, \bibnamefont{and}
  \bibinfo{author}{\bibfnamefont{J.~C.} \bibnamefont{Dyre}},
  \bibinfo{journal}{J. Chem. Phys.} \textbf{\bibinfo{volume}{146}},
  \bibinfo{pages}{154504} (\bibinfo{year}{2017}).

\bibitem[{Gly(1963)}]{GlyProp1963}
\emph{\bibinfo{title}{{Physical Properties of Glycerine and Its Solutions}}}
  (\bibinfo{address}{Glycerine Producers' Association}, \bibinfo{year}{1963}).

\bibitem[{\citenamefont{Donth}(2001)}]{Donth2001}
\bibinfo{author}{\bibfnamefont{E.-J.} \bibnamefont{Donth}},
  \emph{\bibinfo{title}{The Glass Transition: Relaxation Dynamics in Liquids
  and Disordered Materials}}, vol.~\bibinfo{volume}{48} of
  \emph{\bibinfo{series}{Springer Series in Materials Science}}
  (\bibinfo{publisher}{Springer}, \bibinfo{year}{2001}).

\bibitem[{\citenamefont{Angell}(1991)}]{Angell1991}
\bibinfo{author}{\bibfnamefont{C.~A.} \bibnamefont{Angell}},
  \bibinfo{journal}{J. Non-Cryst. Solids} \textbf{\bibinfo{volume}{131}},
  \bibinfo{pages}{13} (\bibinfo{year}{1991}).

\bibitem[{\citenamefont{Lane}(1925)}]{Lane1925}
\bibinfo{author}{\bibfnamefont{L.}~\bibnamefont{Lane}}, \bibinfo{journal}{Ind.
  Eng. Chem} \textbf{\bibinfo{volume}{17}}, \bibinfo{pages}{924}
  (\bibinfo{year}{1925}).

\bibitem[{\citenamefont{Bachler et~al.}(2016)\citenamefont{Bachler,
  Fuentes-Landete, Jahn, Wong, Giovambattista, and Loerting}}]{Bachler2016}
\bibinfo{author}{\bibfnamefont{J.}~\bibnamefont{Bachler}},
  \bibinfo{author}{\bibfnamefont{V.}~\bibnamefont{Fuentes-Landete}},
  \bibinfo{author}{\bibfnamefont{D.~A.} \bibnamefont{Jahn}},
  \bibinfo{author}{\bibfnamefont{J.}~\bibnamefont{Wong}},
  \bibinfo{author}{\bibfnamefont{N.}~\bibnamefont{Giovambattista}},
  \bibnamefont{and} \bibinfo{author}{\bibfnamefont{T.}~\bibnamefont{Loerting}},
  \bibinfo{journal}{Phys. Chem. Chem. Phys.} \textbf{\bibinfo{volume}{18}},
  \bibinfo{pages}{11058} (\bibinfo{year}{2016}).

\bibitem[{\citenamefont{Jakobsen et~al.}(2016)\citenamefont{Jakobsen, Sanz,
  Niss, Hecksher, Pedersen, Rasmussen, Christensen, Olsen, and
  Dyre}}]{Jakobsen2016}
\bibinfo{author}{\bibfnamefont{B.}~\bibnamefont{Jakobsen}},
  \bibinfo{author}{\bibfnamefont{A.}~\bibnamefont{Sanz}},
  \bibinfo{author}{\bibfnamefont{K.}~\bibnamefont{Niss}},
  \bibinfo{author}{\bibfnamefont{T.}~\bibnamefont{Hecksher}},
  \bibinfo{author}{\bibfnamefont{I.~H.} \bibnamefont{Pedersen}},
  \bibinfo{author}{\bibfnamefont{T.}~\bibnamefont{Rasmussen}},
  \bibinfo{author}{\bibfnamefont{T.}~\bibnamefont{Christensen}},
  \bibinfo{author}{\bibfnamefont{N.~B.} \bibnamefont{Olsen}}, \bibnamefont{and}
  \bibinfo{author}{\bibfnamefont{J.~C.} \bibnamefont{Dyre}},
  \bibinfo{journal}{AIP Adv.} \textbf{\bibinfo{volume}{6}},
  \bibinfo{pages}{055019} (\bibinfo{year}{2016}).

\bibitem[{\citenamefont{Harrison}(1976)}]{Harrison1976}
\bibinfo{author}{\bibfnamefont{G.}~\bibnamefont{Harrison}},
  \emph{\bibinfo{title}{The Dynamic Properties of Supercooled Liquids}}
  (\bibinfo{publisher}{Academic press (INC.) London LTD},
  \bibinfo{year}{1976}).

\bibitem[{\citenamefont{Jakobsen et~al.}(2005)\citenamefont{Jakobsen, Niss, and
  Olsen}}]{Jakobsen2005}
\bibinfo{author}{\bibfnamefont{B.}~\bibnamefont{Jakobsen}},
  \bibinfo{author}{\bibfnamefont{K.}~\bibnamefont{Niss}}, \bibnamefont{and}
  \bibinfo{author}{\bibfnamefont{N.~B.} \bibnamefont{Olsen}},
  \bibinfo{journal}{J. Chem. Phys.} \textbf{\bibinfo{volume}{123}},
  \bibinfo{pages}{234511} (\bibinfo{year}{2005}).

\bibitem[{\citenamefont{Roed et~al.}(2015)\citenamefont{Roed, Niss, and
  Jakobsen}}]{Roed2015}
\bibinfo{author}{\bibfnamefont{L.~A.} \bibnamefont{Roed}},
  \bibinfo{author}{\bibfnamefont{K.}~\bibnamefont{Niss}}, \bibnamefont{and}
  \bibinfo{author}{\bibfnamefont{B.}~\bibnamefont{Jakobsen}},
  \bibinfo{journal}{J. Chem. Phys.} \textbf{\bibinfo{volume}{143}}
  (\bibinfo{year}{2015}).

\bibitem[{\citenamefont{B\"ohmer et~al.}(1993)\citenamefont{B\"ohmer, Ngai,
  Angell, and Plazek}}]{Bohmer1993}
\bibinfo{author}{\bibfnamefont{R.}~\bibnamefont{B\"ohmer}},
  \bibinfo{author}{\bibfnamefont{K.~L.} \bibnamefont{Ngai}},
  \bibinfo{author}{\bibfnamefont{C.~A.} \bibnamefont{Angell}},
  \bibnamefont{and} \bibinfo{author}{\bibfnamefont{D.~J.}
  \bibnamefont{Plazek}}, \bibinfo{journal}{J. Chem. Phys.}
  \textbf{\bibinfo{volume}{99}}, \bibinfo{pages}{4201} (\bibinfo{year}{1993}).

\bibitem[{\citenamefont{Niss et~al.}(2005)\citenamefont{Niss, Jakobsen, and
  Olsen}}]{Niss2005}
\bibinfo{author}{\bibfnamefont{K.}~\bibnamefont{Niss}},
  \bibinfo{author}{\bibfnamefont{B.}~\bibnamefont{Jakobsen}}, \bibnamefont{and}
  \bibinfo{author}{\bibfnamefont{N.~B.} \bibnamefont{Olsen}},
  \bibinfo{journal}{J. Chem. Phys.} \textbf{\bibinfo{volume}{123}},
  \bibinfo{pages}{234510} (\bibinfo{year}{2005}).

\bibitem[{\citenamefont{Jakobsen et~al.}(2011)\citenamefont{Jakobsen, Niss,
  Maggi, Olsen, Christensen, and Dyre}}]{Jakobsen2011}
\bibinfo{author}{\bibfnamefont{B.}~\bibnamefont{Jakobsen}},
  \bibinfo{author}{\bibfnamefont{K.}~\bibnamefont{Niss}},
  \bibinfo{author}{\bibfnamefont{C.}~\bibnamefont{Maggi}},
  \bibinfo{author}{\bibfnamefont{N.~B.} \bibnamefont{Olsen}},
  \bibinfo{author}{\bibfnamefont{T.}~\bibnamefont{Christensen}},
  \bibnamefont{and} \bibinfo{author}{\bibfnamefont{J.~C.} \bibnamefont{Dyre}},
  \bibinfo{journal}{J. Non-Cryst. Solids} \textbf{\bibinfo{volume}{357}},
  \bibinfo{pages}{267} (\bibinfo{year}{2011}).

\bibitem[{\citenamefont{Kremer and Sch\"{o}nhals}(2003)}]{Kremer2003}
\bibinfo{editor}{\bibfnamefont{F.}~\bibnamefont{Kremer}} \bibnamefont{and}
  \bibinfo{editor}{\bibfnamefont{A.}~\bibnamefont{Sch\"{o}nhals}}, eds.,
  \emph{\bibinfo{title}{Broadband dielectric spectroscopy}}
  (\bibinfo{publisher}{Springe}, \bibinfo{address}{Berlin},
  \bibinfo{year}{2003}).

\bibitem[{\citenamefont{Hofmann et~al.}(2015)\citenamefont{Hofmann, Gainaru,
  Cetinkaya, Valiullin, Fatkullin, and R\"ossler}}]{Hofman2015}
\bibinfo{author}{\bibfnamefont{M.}~\bibnamefont{Hofmann}},
  \bibinfo{author}{\bibfnamefont{C.}~\bibnamefont{Gainaru}},
  \bibinfo{author}{\bibfnamefont{B.}~\bibnamefont{Cetinkaya}},
  \bibinfo{author}{\bibfnamefont{R.}~\bibnamefont{Valiullin}},
  \bibinfo{author}{\bibfnamefont{N.}~\bibnamefont{Fatkullin}},
  \bibnamefont{and} \bibinfo{author}{\bibfnamefont{E.~A.}
  \bibnamefont{R\"ossler}}, \bibinfo{journal}{Macromolecules}
  \textbf{\bibinfo{volume}{48}}, \bibinfo{pages}{7521} (\bibinfo{year}{2015}).

\bibitem[{\citenamefont{Egorov et~al.}(2013)\citenamefont{Egorov, Makarov, and
  Kolker}}]{Egorov2013}
\bibinfo{author}{\bibfnamefont{G.~I.} \bibnamefont{Egorov}},
  \bibinfo{author}{\bibfnamefont{D.~M.} \bibnamefont{Makarov}},
  \bibnamefont{and} \bibinfo{author}{\bibfnamefont{A.~M.}
  \bibnamefont{Kolker}}, \bibinfo{journal}{Thermochim. Acta}
  \textbf{\bibinfo{volume}{570}}, \bibinfo{pages}{16} (\bibinfo{year}{2013}).

\bibitem[{\citenamefont{Egorov et~al.}(2011)\citenamefont{Egorov, Lyubartsev,
  and Laaksonen}}]{Egorov2011}
\bibinfo{author}{\bibfnamefont{A.~V.} \bibnamefont{Egorov}},
  \bibinfo{author}{\bibfnamefont{A.~P.} \bibnamefont{Lyubartsev}},
  \bibnamefont{and}
  \bibinfo{author}{\bibfnamefont{A.}~\bibnamefont{Laaksonen}},
  \bibinfo{journal}{J. Phys. Chem. B} \textbf{\bibinfo{volume}{115}},
  \bibinfo{pages}{14572} (\bibinfo{year}{2011}).

\bibitem[{\citenamefont{Towey et~al.}(2011{\natexlab{a}})\citenamefont{Towey,
  Soper, and Dougan}}]{Towey2011}
\bibinfo{author}{\bibfnamefont{J.~J.} \bibnamefont{Towey}},
  \bibinfo{author}{\bibfnamefont{A.~K.} \bibnamefont{Soper}}, \bibnamefont{and}
  \bibinfo{author}{\bibfnamefont{L.}~\bibnamefont{Dougan}},
  \bibinfo{journal}{Phys. Chem. Chem. Phys.} \textbf{\bibinfo{volume}{13}},
  \bibinfo{pages}{9397} (\bibinfo{year}{2011}{\natexlab{a}}).

\bibitem[{\citenamefont{Towey et~al.}(2011{\natexlab{b}})\citenamefont{Towey,
  Soper, and Dougan}}]{Towey2011a}
\bibinfo{author}{\bibfnamefont{J.~J.} \bibnamefont{Towey}},
  \bibinfo{author}{\bibfnamefont{A.~K.} \bibnamefont{Soper}}, \bibnamefont{and}
  \bibinfo{author}{\bibfnamefont{L.}~\bibnamefont{Dougan}},
  \bibinfo{journal}{J. Phys. Chem. B} \textbf{\bibinfo{volume}{115}},
  \bibinfo{pages}{7799} (\bibinfo{year}{2011}{\natexlab{b}}).

\bibitem[{\citenamefont{Towey and Dougan}(2012)}]{Towey2012}
\bibinfo{author}{\bibfnamefont{J.~J.} \bibnamefont{Towey}} \bibnamefont{and}
  \bibinfo{author}{\bibfnamefont{L.}~\bibnamefont{Dougan}},
  \bibinfo{journal}{J. Phys. Chem. B} \textbf{\bibinfo{volume}{116}},
  \bibinfo{pages}{1633} (\bibinfo{year}{2012}).

\bibitem[{\citenamefont{Towey et~al.}(2012)\citenamefont{Towey, Soper, and
  Dougan}}]{Towey2012b}
\bibinfo{author}{\bibfnamefont{J.~J.} \bibnamefont{Towey}},
  \bibinfo{author}{\bibfnamefont{A.~K.} \bibnamefont{Soper}}, \bibnamefont{and}
  \bibinfo{author}{\bibfnamefont{L.}~\bibnamefont{Dougan}},
  \bibinfo{journal}{J. Phys. Chem. B} \textbf{\bibinfo{volume}{116}},
  \bibinfo{pages}{13898} (\bibinfo{year}{2012}).

\bibitem[{\citenamefont{Towey et~al.}(2013)\citenamefont{Towey, Soper, and
  Dougan}}]{Towey2013}
\bibinfo{author}{\bibfnamefont{J.~J.} \bibnamefont{Towey}},
  \bibinfo{author}{\bibfnamefont{A.~K.} \bibnamefont{Soper}}, \bibnamefont{and}
  \bibinfo{author}{\bibfnamefont{L.}~\bibnamefont{Dougan}},
  \bibinfo{journal}{Faraday Discuss.} \textbf{\bibinfo{volume}{167}},
  \bibinfo{pages}{159} (\bibinfo{year}{2013}).

\bibitem[{\citenamefont{Artola et~al.}(2013)\citenamefont{Artola, Raihane,
  Crauste-Thibierge, Merlet, Emo, Alba-Simionesco, and Rousseau}}]{Artola2013}
\bibinfo{author}{\bibfnamefont{P.~A.} \bibnamefont{Artola}},
  \bibinfo{author}{\bibfnamefont{A.}~\bibnamefont{Raihane}},
  \bibinfo{author}{\bibfnamefont{C.}~\bibnamefont{Crauste-Thibierge}},
  \bibinfo{author}{\bibfnamefont{D.}~\bibnamefont{Merlet}},
  \bibinfo{author}{\bibfnamefont{M.}~\bibnamefont{Emo}},
  \bibinfo{author}{\bibfnamefont{C.}~\bibnamefont{Alba-Simionesco}},
  \bibnamefont{and} \bibinfo{author}{\bibfnamefont{B.}~\bibnamefont{Rousseau}},
  \bibinfo{journal}{J. Phys. Chem. B} \textbf{\bibinfo{volume}{117}},
  \bibinfo{pages}{9718} (\bibinfo{year}{2013}), \bibinfo{note}{pMID: 23937163}.

\end{thebibliography}

\end{document}